\documentclass[aps, twocolumn, superscriptaddress, amsmath, amssymb, floatfix, prb,reprint, longbibliography,footnoteinbib, notitlepage]{revtex4-2} 
\usepackage{amsbsy}
\usepackage{amsfonts}
\usepackage{graphicx}
\usepackage{bm}
\usepackage{bbm}
\usepackage{color}
\definecolor{prlblue}{rgb}{0.18,0.19,0.57}
\usepackage{multirow}
\usepackage[colorlinks]{hyperref}
\usepackage{cleveref}
\usepackage[T1]{fontenc}
\usepackage{ifthen}
\usepackage{tikz}
\usetikzlibrary{positioning}
\usepackage{multirow}
\usepackage{braket}

\usetikzlibrary{patterns.meta} 
\usetikzlibrary{calc}

\renewcommand{\i}{\mathrm{i}}
\newcommand{\x}{\bm{x}}
\newcommand{\g}{\bm{g}}
\renewcommand{\r}{\bm{r}}
\newcommand{\R}{\bm{R}}
\renewcommand{\d}{\bm{d}}
\renewcommand{\k}{\bm{k}}
\newcommand{\rom}[1]{\uppercase\expandafter{\romannumeral #1\relax}}

\newcommand{\be}{\begin{equation}}
\newcommand{\ee}{\end{equation}}

\begin{document}
\graphicspath{{figures/}}

\title{Moiré Engineering of Square-Lattice Hubbard Models \\ for Copper- and Iron-Based Superconductors}

\author{Toshikaze Kariyado}
\affiliation{International Center for Materials Nanoarchitectonics (WPI-MANA), 
National Institute for Materials Science, Tsukuba 305-0044, Japan}

\author{Yusuf Wicaksono}
\affiliation{International Center for Materials Nanoarchitectonics (WPI-MANA), 
National Institute for Materials Science, Tsukuba 305-0044, Japan}

\author{Ashvin Vishwanath}
\affiliation{Department of Physics, Harvard University, Cambridge, MA 02138, USA}

\author{Pavel A. Volkov}
\affiliation{Department of Physics, University of Connecticut, Storrs, Connecticut 06269, USA}

\author{Zhu-Xi Luo}
\affiliation{School of Physics, Georgia Institute of Technology, Atlanta, Georgia 30332, USA}

\date{\today}

\begin{abstract}
We propose twisted $\Gamma$-valley square-lattice homobilayers, exemplified by ZnF$_2$, as a tunable reference platform to emulate square lattice Hubbard-type models relevant to high-temperature superconductors. With model parameters derived from first-principles calculations, the lowest moiré band realizes a single-orbital Hubbard model underlying cuprates, while the next two realize a $p_x, p_y$ model with the minimal symmetry structure of pnictide $d_{xz}$, $d_{yz}$ Hamiltonians. Hartree-Fock analysis further predicts an antiferro-orbital ferromagnetic insulator at quarter filling.
\end{abstract}

\maketitle

\tableofcontents

\section{Introduction} 
Moiré heterostructures of van der Waals (vdW) materials have proven to be powerful and versatile platforms to simulate important models in condensed matter physics as well as to generate novel strongly correlated quantum phenomena. 
While rich correlated physics has been revealed in moiré systems — especially superconductivity \cite{cao2018unconventional,doi:10.1126/science.aav1910,lu2019superconductors,park2021tunable,doi:10.1126/science.abg0399,park2022robust,chen2019signatures,zhou2021superconductivity,kim2022evidence,xia2024unconventionalsuperconductivitytwistedbilayer,guo2024superconductivitytwistedbilayerwse2} — research has so far focused almost exclusively on hexagonal lattices. In contrast, the canonical high-temperature superconductors—cuprates \cite{proust2019remarkable}, iron-based superconductors \cite{fernandes2022iron,si2016high}, and most recently the nickelate families \cite{zhou2025ambient} — all share a square-lattice foundation. These materials exhibit unconventional superconductivity intertwined with strongly correlated phenomena such as charge and spin stripes, pseudogap behavior, and exotic metallicity—features that remain among the most profound challenges in condensed matter physics. Their intrinsic chemical complexity and limited tunability have long obscured the underlying microscopic mechanisms. 
Given the exceptional tunability of vdW heterostructures, square-lattice moiré platforms provide a promising route to realize the effective Hubbard-type models believed to underlie high-$T_c$ superconductors, offering a controlled reference system to identify where real materials deviate from this idealized starting point and to uncover correlated phenomena beyond it. 

Most theoretical studies of twisted square lattices emphasize single-particle physics and band engineering \cite{PhysRevB.105.165422,PhysRevB.104.035136,PhysRevResearch.1.033076,PhysRevResearch.4.043151,bao2026moireferroelectricitydrivenbandengineering}, leaving strong correlations largely unexplored. Experimental attempts to realize these systems include engineering preexisting monolayer superconductors \cite{can2021high,doi:10.1126/science.abl8371,PhysRevB.105.L201102,PhysRevB.106.104505,PhysRevLett.130.186001,10.21468/SciPostPhys.15.3.081} and utilizing analog cold-atom \cite{cirac2019,salamon2020,meng2023atomic}, optical \cite{fu2020optical,wang2020localization,
lou2021}, and acoustic \cite{phon2022} platforms. However, vdW heterostructures built from normal parent layers uniquely offer the requisite interaction strength and in situ tunability \cite{yang2025engineeringhubbardmodelsgated} to engineer strongly correlated Hubbard models and simulate high-$T_c$ physics.

Earlier attempts by some of the authors to simulate high-$T_c$ physics in moiré systems focused on the $M$-valley \cite{PhysRevLett.134.236503,ToshiLuo} {partially motivated by the symmetry-enforced moir\'e interference independent of microscopic details \cite{PhysRevResearch.1.033076}. However,} the scarcity of experimentally viable materials has limited their realization{: for example, in the database  of computationally exfoliable square lattice materials~\cite{twistable_dictionary}, only $2$ $M$-valley candidates were identified, as opposed to the $23$ $\Gamma$-valley candidates. The two constructions differ structurally: at small twist angles, $M$-valley bilayers possess an emergent symmetry that forbids nearest-neighbor hopping $t$ on the moir\'e superlattice, so that the next-nearest-neighbor hopping $t'$ dominates and a displacement field is needed to couple the two sublattices and tune $t'/t$ \cite{PhysRevLett.134.236503}. In contrast, no such symmetry exists at $\Gamma$ which opens up the possibility for a single-orbital Hubbard model with nearest-neighbor hopping. Nevertheless,} the relevant literature on $\Gamma$-valley square-lattice moir\'e heterostructures is sparse \cite{xu2024engineering2dsquarelattice,10.21468/SciPostPhys.15.3.081} and a general framework is yet unavailable.

This work will close this gap through a systematic study of the continuum model and the resultant effective Hubbard-type model on the moiré superlattice, and further examine ZnF$_2$ as a promising candidate. Interestingly, the first moir\'e band is found to realize the single-orbital square lattice Hubbard model which is widely believed to capture essential cuprate physics \cite{proust2019remarkable}, while the following two moi\'re bands are found to realize two-orbital square lattice Hubbard models  closely related to \footnote{Although these models use different symmetry assignments  from the microscopically accurate models \cite{Lee_2008,Kuroki_2008},  leading to different Fermi surface topology in a symmetry extended Brillouin zone, they are still believed to  capture key qualitative aspects of the  relevant physics of the pnictides \cite{Wang_2009,wang2008numerical}.} the effective $d_{xz}$, $d_{yz}$-models proposed for iron pnictides \cite{PhysRevB.77.220503,PhysRevLett.101.237004,PhysRevB.79.134502}. The moiré realization is demonstrated in the general analytical analysis and further supported by first-principle calculations in the twisted homobilayer of ZnF$_2$. We also present new Hartree-Fock calculations for the two-orbital model at quarter filling, which could be of separate interest.

\section{Gamma point material} 
For a monolayer with a $\Gamma$-valley extremum, the effective twisted bilayer Hamiltonian is:
\be
H=
\begin{pmatrix}
H_0(-\i \bm{\nabla}_{+\theta/2} )+V(\bm{x}) & T(\bm{x}) \\
T(\bm{x})^* & H_0(-\i \bm{\nabla}_{-\theta/2} )+V(\bm{x}) \\
\end{pmatrix},
\label{eq:TB_Hamiltonian}
\ee
where $H_0(\mathbf{k}) = \hbar^2\mathbf{k}^2/2m_*$ with $m^*$ being the effective mass. Spin-orbit coupling is neglected, which vanishes at the $\Gamma$-point, and is absent in presence of inversion (such as in the case of ZnF$_2$ studied below) at the single-layer level. 
Moir\'e periodicity ensures that $V(\x)$ to be periodic and time reversal symmetry constrains it to be real. Fourfold rotation $C_{4z}$ require that $V(\x)=V(C_{4z}\x)$, and reflection symmetries with respect to the $x$-axis requires $V(x,y)=V(-x,y)$. Similar constraints apply to the interlayer hopping $T(\x)$ as well. To the lowest harmonics, this yields
\be
\begin{split}
& V(\x)=V_1 \left[ \cos (\g_1\cdot \x) + \cos (\g_2\cdot \x)\right],\\
& T(\x)=T_0 + T_1 \left[ \cos (\g_1\cdot \x) + \cos (\g_2\cdot \x) \right],
\label{eq:leading}
\end{split}
\ee
where $\g_i$ are the moir\'e reciprocal lattice vectors satisfying $g=|\g_i|=2\pi/a_m$, with moir\'e lattice constant $a_m\approx a_0/\theta$. At small twist angles, we neglect the $\pm \theta/2$ relative rotation in the intralayer kinetic energy $H_0$. The Hamiltonian then diagonalizes in the $\sigma^x$ basis and separates into  $H_{\pm}=H_{\pm,x}+H_{\pm,y},$ where $
H_{\pm, x_i} = H_0 (-\i {\nabla}_i)\pm \frac{1}{2} T_0+(V_1\pm T_1) \cos (g_i x_i).
$ 
This reduces the Schrödinger's equation to decoupled 1D Mathieu equations, whose solutions are the Mathieu functions as reviewed in {Appendix \ref{app:Matt}}.

\subsection{Hopping parameters} 
For the lowest $s$-orbital band as $\theta\rightarrow 0$, nearest-neighbor hopping dominates with dispersion $\varepsilon_s (\k)=E_s-2t (\cos k_x + \cos k_y)$. 
The hopping amplitude can thus be obtained from $t\propto [\varepsilon_s (M_m) - \varepsilon_s ( \Gamma_m) ]/8$ and the asymptotic forms of Mathieu functions, \be
t \sim \frac{\hbar^2\pi^2}{m_*a_m^2} \frac{8\sqrt{2}}{\sqrt{\pi}} B^{\frac{3}{4}} e^{-4\sqrt{B}},
\label{eq:t_s}
\ee
where $B=\frac{m_* a_m^2}{\hbar^2\pi^2}(V_1+T_1)$ $\gg 1$ scales as $a_m^2$ and goes to infinity in the small angle asymptotic limit. Eqn. \eqref{eq:t_s} is consistent with optical lattice results \cite{Arzamasovs_2017}. The next two lowest bands correspond to the $p_x, p_y$ orbitals with dispersions $\varepsilon_{p_x} = E_{p}-2t_{\sigma} \cos k_x - 2t_{\pi} \cos k_y$, and $\varepsilon_{p_y} =E_{p} -2t_{\pi} \cos k_x - 2t_{\sigma} \cos k_y$, respectively. For concreteness we will assume $t_{\sigma}<t_{\pi}$. The two hopping parameters can again be extracted from the dispersions and Mathieu functions, leading to \be 
t_{\pi}\sim t,\quad t_{\sigma}\sim - 16B^{1/2}t_{\pi}.
\label{eq:t_p}
\ee 
Notably, $t_{\sigma}$ and $t_{\pi}$ have opposite signs and distinct scalings. Higher orbital bands ($d$, $f$, etc.) can be similarly modeled and accessed experimentally via gating.

\subsection{Coulomb interactions} 
To further estimate the Coulomb interactions, we expand the periodic potentials in $H_{\pm,i}$ to quadratic order near their minima. This yields
1D harmonic oscillators \footnote{We remark that while the harmonic oscillator approximation is powerful in estimating the Coulomb interactions in the asymptotic limit, it gives wrong pre-exponential as well as exponential factors \cite{Arzamasovs_2017} for hopping parameters, which is why we chose to use the Mathieu treatment instead. } with characteristic frequencies
$\omega_{\pm} =g\sqrt{|V_1\pm T_1|/m_*}$. Defining $\alpha_{\pm}=m_*\omega_{\pm}/\hbar$,
the lowest-energy wavefunctions are 
$\psi_{s} = \sqrt{\alpha/\pi} e^{-\alpha (x^2+y^2)/2},$ and $\psi_{p_i}=\sqrt{2\alpha}x_i\psi_s$. The density-density interaction between $s$-orbitals at separation $\bm{d}$ then integrates to $U_{ss}(\bm{d})=\frac{e^2}{4\pi \epsilon} \frac{\sqrt{\pi \alpha}}{2\sqrt{2}} I_0(\alpha d^2/4) e^{-\alpha d^2/4}$, where $I$ is the modified Bessel function of the first kind.  This decays as $\sim 1/d$ as expected. At $d=0$, the onsite repulsion is \be U_{ss}(0)=\frac{e^2}{4\pi \epsilon} \frac{\sqrt{\pi \alpha}}{2\sqrt{2}}.\label{eq:Uss}\ee

Consistent results were obtained in ref. \cite{PhysRevLett.134.236503} in a different context. For the $p$-orbitals, analogous integration  yields onsite repulsions $U_{p_xp_x}(0) = \frac{25}{32} U_{ss}(0)$ and $U_{p_xp_y}(0)=\frac{19}{32} U_{ss}(0)$, alongside a weaker Hund's coupling $J_{p_xp_y}(0)=\frac{3}{16} U_{ss}(0)$, see {Appendix \ref{app:Gaussian} for details.}

\subsection{Antiferro-Orbital Order}
\begin{figure}[htbp]
\begin{tikzpicture}[
    scale=0.85, 
    orbital/.style={blue, thick, fill=white}, 
    lattice/.style={gray!60, thin}, 
    pics/px/.style={
        code={
            \draw[orbital] (0.2,0) ellipse (0.2 and 0.12);
            \draw[orbital] (-0.2,0) ellipse (0.2 and 0.12);
        }
    },
    pics/py/.style={
        code={
            \draw[orbital] (0,0.2) ellipse (0.12 and 0.2);
            \draw[orbital] (0,-0.2) ellipse (0.12 and 0.2);
        }
    }
]
\begin{scope}[local bounding box=fig_a]
    \draw[lattice] (-0.5, -0.5) grid (2.5, 2.5);
    \foreach \x in {0,1,2} {
        \foreach \y in {0,1,2} {
            \pic at (\x, \y) {py};
        }
    }
\end{scope}
\node[below=5pt of fig_a] {(a)};
\begin{scope}[xshift=3.5cm, local bounding box=fig_b]
    \draw[lattice] (-0.5, -0.5) grid (2.5, 2.5);
    \foreach \x in {0,1,2} {
        \foreach \y in {0,1,2} {
            \ifnum\y=1 
                \pic at (\x, \y) {py};
            \else
                \pic at (\x, \y) {px};
            \fi
        }
    }
\end{scope}
\node[below=5pt of fig_b] {(b)};
\begin{scope}[xshift=7cm, local bounding box=fig_c]
    \draw[lattice] (-0.5, -0.5) grid (2.5, 2.5);
    \foreach \x in {0,1,2} {
        \foreach \y in {0,1,2} { \pgfmathparse{mod(\x+\y,2)}
            \ifdim\pgfmathresult pt=0 pt
                \pic at (\x, \y) {px}; 
            \else
                \pic at (\x, \y) {py}; 
            \fi
        }
    }
\end{scope}
\node[below=5pt of fig_c] {(c)};
\end{tikzpicture}
\caption{Orbital configurations for (a) ferro-orbital order, (b) stripe-orbital order, and (c) antiferro-orbital order. They have energies $E^{(a)} =U_A + U_C,$  $E^{(c)} = 2U_B,$ and
$E^{(b)} = U_B + (U_A+U_C)/2$  respectively, and satisfies $E^{(a)}>E^{(b)}>E^{(c)}$.}
\label{fig:orbitals}
\end{figure}
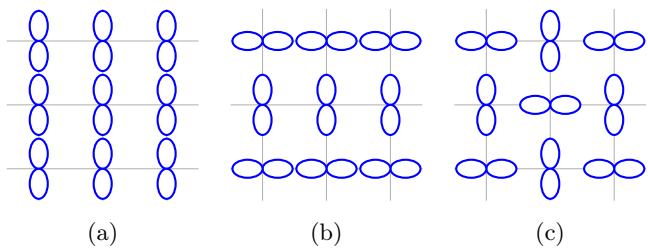

At quarter filling of the $p$-orbitals, strong on-site Coulomb interactions enforce single  occupancy. Because inter-site hopping decays exponentially with the moir\'e period $a_m$ while Coulomb interactions decay algebraically, interactions overwhelm the kinetic energy. Consequently, classical electrostatics dictate the orbital configuration, while exponentially suppressed superexchange leaves the spin degrees of freedom macroscopically degenerate at zeroth order. To determine the orbital order, we compare the energies of the ferro-, stripe-, and antiferro-orbital configurations (Fig. \ref{fig:orbitals}). Defining the nearest-neighbor repulsions $U_{A} \equiv U_{p_xp_x}(a_m,0) = U_{p_yp_y}(0,a_m)$, $U_{B} \equiv U_{p_xp_y}(a_m,0) =U_{p_yp_x}(0,a_m)$, and  $U_{C} \equiv U_{p_xp_x}(0,a_m) = U_{p_yp_y}(a_m,0)$, their asymptotic expansions yield (detailed derivations in {Appendix \ref{app:Gaussian}}):
\be
U_A+U_C-2U_B\sim \frac{57}{8\alpha^2 a_m^5}.
\label{eq:AFO}
\ee
This positive quantity determines the energy hierarchy between the orbital arrangements:  antiferro-orbital ordering is favored over orbital stripes, and the latter is favored over ferro-orbital ordering, see fig.\ref{fig:orbitals}. Therefore, the antiferro-orbital order is always favored at small twisting angles. 

\section{Twisted bilayer {ZnF}$_2$} 
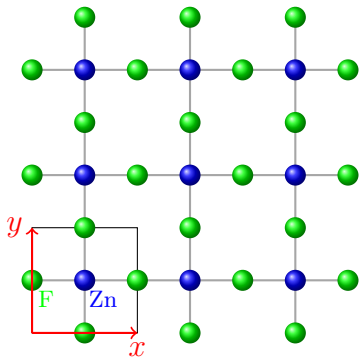
\begin{figure}[htbp]
\centering
\begin{tikzpicture}[scale=0.6]
\foreach \i in {1,3,5}{
  \foreach \j in {1,3,5}{
    \ifnum\i<6
      \draw[gray!70,thick] (\i-1,\j,0) -- (\i+1,\j,0);
    \fi
    \ifnum\j<6
      \draw[gray!70,thick] (\i,\j-1,0) -- (\i,\j+1,0);
    \fi
  }
}    
\draw[black] (0,0,0) -- (2,0,0) -- (2,2,0) -- (0,2,0) -- cycle;
\foreach \i in {0,2,4,6}{ 
  \ifnum \i<6
    \shade[ball color=green] (\i+1,6) circle (0.2);  
  \fi
  
  \foreach \j in {0,2,4,6}{  
    \ifnum\i<6
     \ifnum\j<6
        \shade[ball color=green] (\i+1,\j,0) circle (0.2);
        \shade[ball color=green] (\i,\j+1,0) circle (0.2);      
        \shade[ball color=blue] (\i+1,\j+1,0) circle (0.2);
      \fi
    \fi
  }
}    
\shade[ball color=green] (6,1) circle (0.2);  
\shade[ball color=green] (6,3) circle (0.2);  
\shade[ball color=green] (6,5) circle (0.2);  
\draw[->,thick,red]   (0,0,0) -- (2,0,0) node[below] {\large{$x$}};
\draw[->,thick,red] (0,0,0) -- (0,2,0) node[left]  {\large{$y$}};
\node[blue] at (1.35,0.65) {{Zn}};
\node[green] at (0.27,0.65) {{F}};
\end{tikzpicture}
\caption{Crystal structure of monolayer ZnF$_2$.}
\label{fig:uc}
\end{figure}

ZnF${}_2$ is a two-dimensional material on square lattice. It was identified in first-principle calculations \cite{twistable_dictionary,doi:10.1021/acsnano.2c11510} as computationally stable and twistable, with the 3D parent $[NH_4]_2 ZnF_4$  \cite{parent}. The crystal structure is shown in fig. \ref{fig:uc} with space group $P4/mmm$ and lattice constant $3.87$ Angstrom. We will focus on the conduction band minimum which resides at the $\Gamma$ point with dominant orbital being Zn $4s$. In contrast, the valence band top is at the $M$-point and the relevant orbitals are the Zn $d_{x^2-y^2}$ and F $p_x, p_y$ orbitals, which are all in-plane and produce only weak moir\'e effect (bilayer band structures for different relative stackings do not show clear moir\'e modulations). The monolayer band structure and orbital information are listed in {Appendix \ref{app:DFT}}. 

\subsection{Modeling from first-principle calculations} 
\begin{figure}[htbp]
\centering
\includegraphics[scale=0.285]{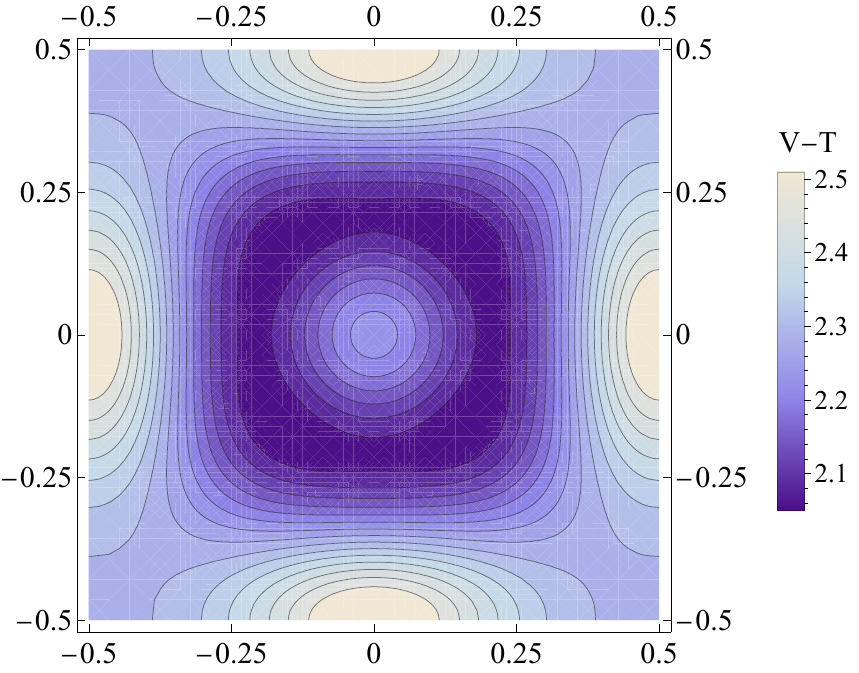}
\includegraphics[trim={0 0 0 0},clip, scale=0.23]{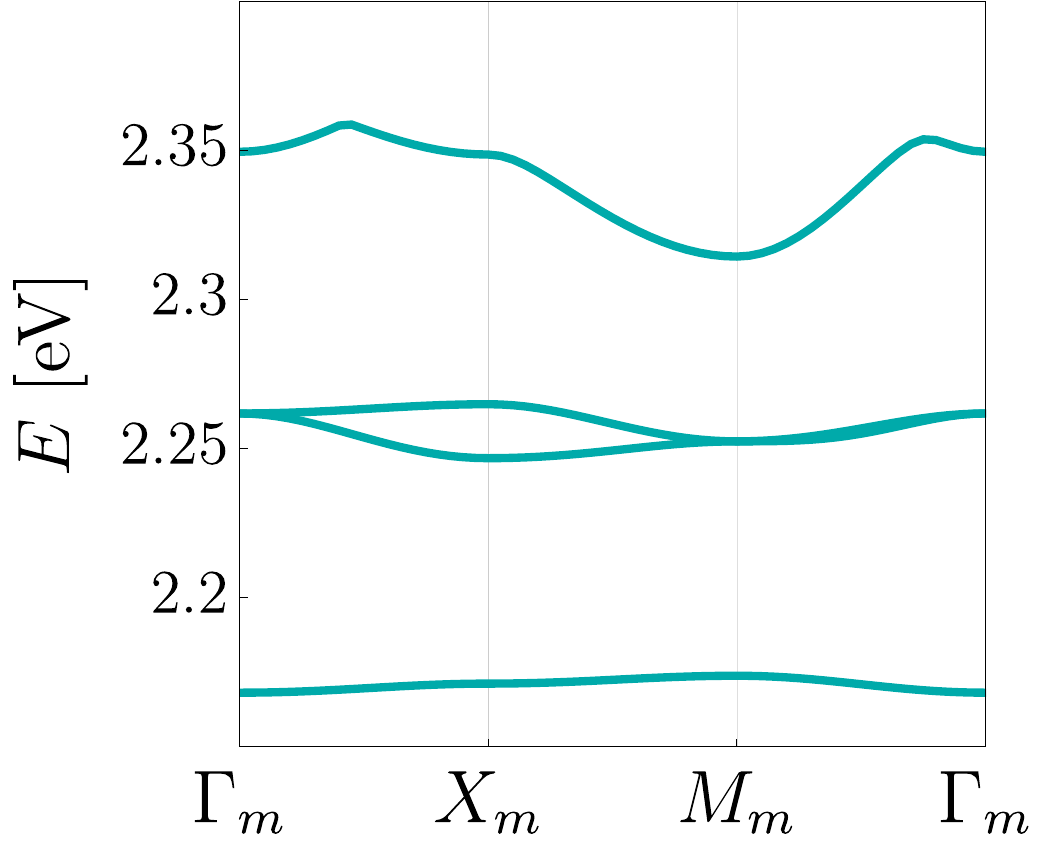}
\caption{
Left: Real-space profiles for moiré potential $V-T$ of twisted bilayer ZnF$_2$, plotted as a function of normalized coordinates within the moiré unit cell in eV. Right: Moir\'e band structures at twisting angle $\theta=2^{\circ}$.
}
\label{fig:corrugation_compare}
\end{figure}

To fit the parameters in the continuum Hamiltonian \eqref{eq:TB_Hamiltonian} from first-principle calculations, we approximate the local twisted bilayer environment by untwisted bilayers with different relative constant shifts. For each in-plane shift $\bm{\tau}$, we vary the total bilayer energy as a function of interlayer distances to find the optimal $d_z (\bm{\tau})$ associated with the lowest total energy. The band structure calculation for the untwisted bilayer is then performed with this optimal $d_z (\bm{\tau})$. 
The difference between the conduction band minima for the lowest and second lowest conduction bands is twice $T[\bm{\tau}(\bm{x})]$, and the average of these two energies lead to $V[\bm{\tau}(\x)]$. We keep track of the first 5 harmonics, with the detailed coefficients presented in {Appendix \ref{app:DFT}}. In the left panel of fig. \ref{fig:corrugation_compare}, we plot the combination of the moiré potential $V(\bm{x})-T(\bm{x})$ responsible for the band structures.  

\begin{figure}[hbtp]
\centering
\includegraphics[trim={0cm 0 0 0},clip, scale=0.25]{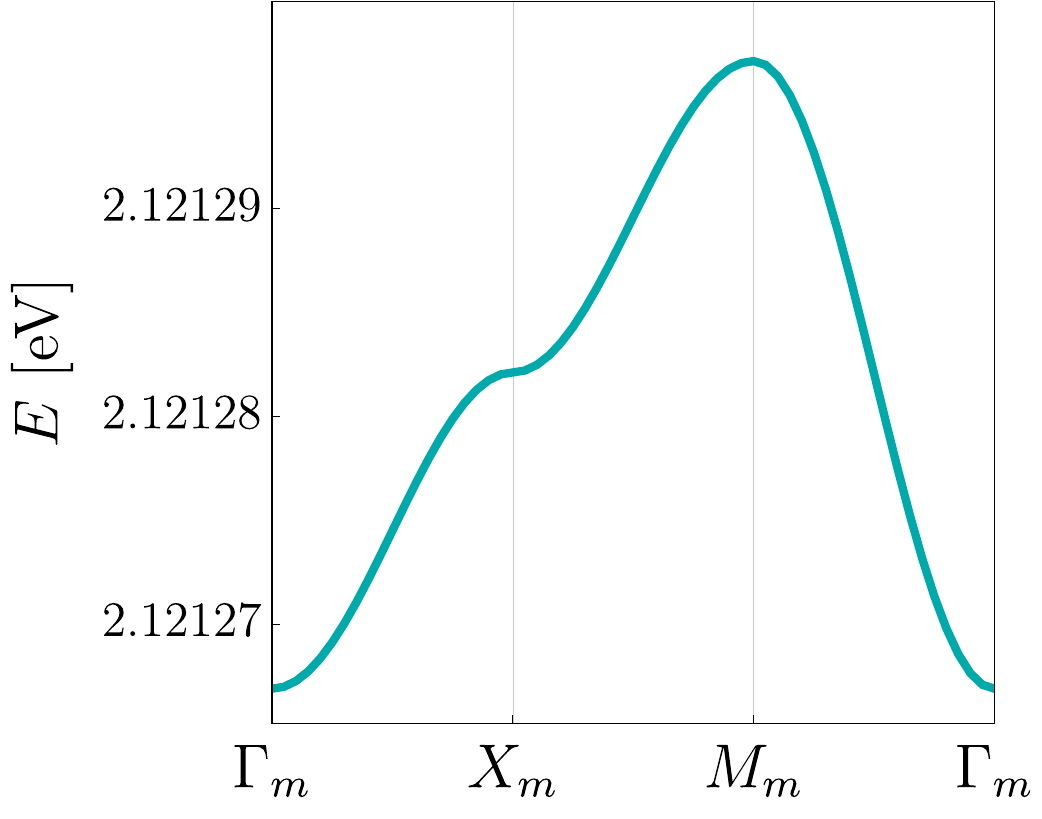}
\includegraphics[trim={1.6cm 0 0 0},clip, scale=0.25]{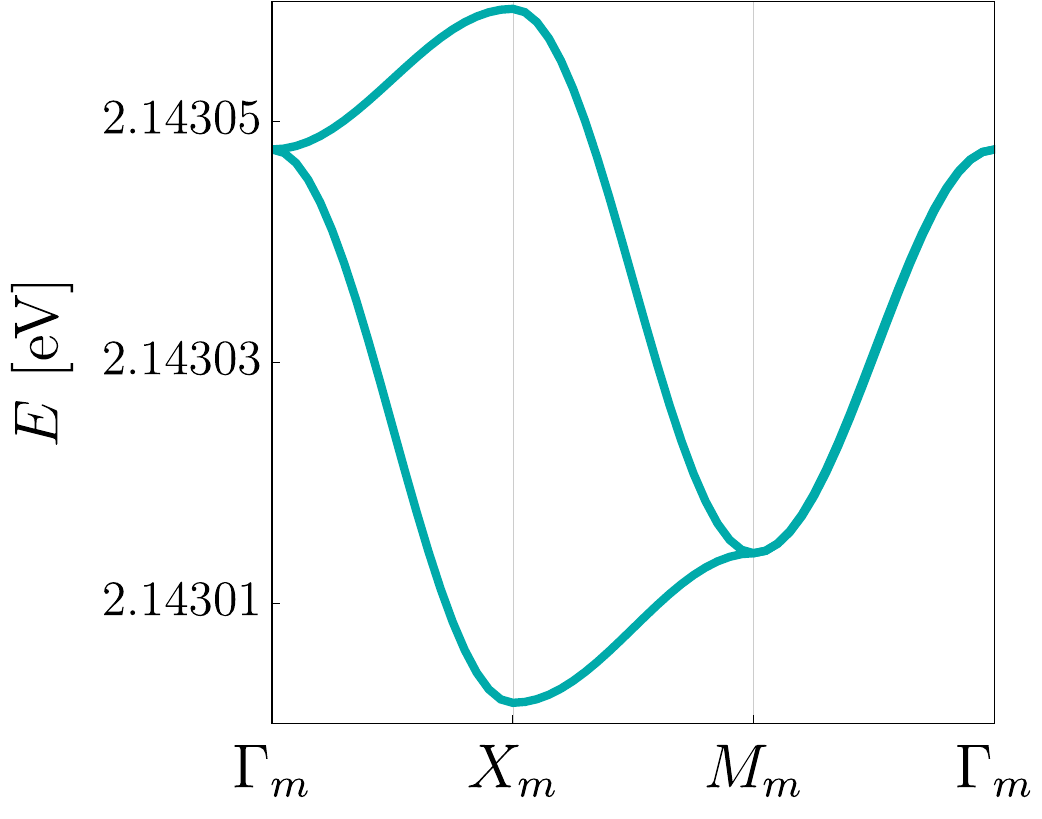}
\caption{Band structures at $\theta=1^{\circ}$. Left: the lowest-lying band corresponding to moir\'e $s$-orbital. Right: the next two low-lying bands corresponding to the moir\'e $p_x,$ $p_y$ orbitals.}
\label{fig:separate_bands}
\end{figure}

\smallskip
\noindent\textit{Band structures.} Fig. \ref{fig:corrugation_compare} shows the band structures of the twisted bilayer model at $\theta=2^{\circ}$, and comments on higher twisting angle information can be found in {Appendix \ref{app:higher_angle}}. The gap between the lowest and neighboring moir\'e conduction bands is of order $50$ meV and therefore the higher bands can easily be accessed via gating. 
Zooming in near the lowest moir\'e conduction band and the next two low-lying bands, we see in fig. \ref{fig:separate_bands} that they manifest the $s$- and $p_x$, $p_y$ orbital physics consistent with the previous analysis. 

\subsection{Hubbard parameters} 
\begin{widetext}
\begin{figure*}[htb]
\includegraphics[scale=0.85]{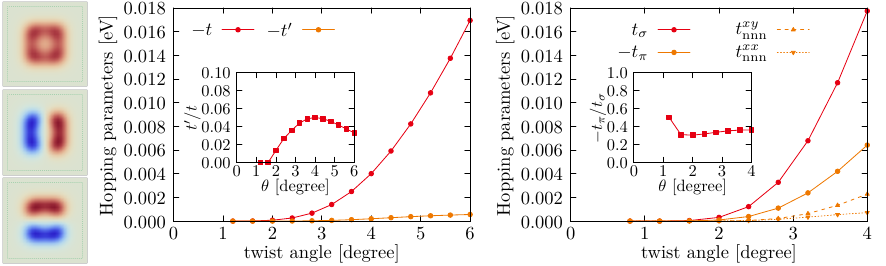}
\caption{
Left: real-space distributions of the Wannier functions for the $s$, $p_y$ and $p_x$ orbitals. Middle: $s$-orbital hopping parameters for nearest neighbor $t$ and next nearest neighbor $t'$. Inset: Ratio $t'/t$ as a function of twisting angle $\theta$. Right: $p_x$, $p_y$- orbital hopping parameters for nearest neighbor $t_{\sigma}, t_{\pi}$ and next nearest neighbor $t^{xy}$ and $t^{xx}$. Inset: Ratio $-t_{\pi}/t_{\sigma}$ as a function of $\theta$. }
\label{fig:hopping_fit}
\end{figure*}
\end{widetext}

\begin{figure}[hbt]
\includegraphics[scale=0.78,trim={0 0 5.5cm 0 }, clip]{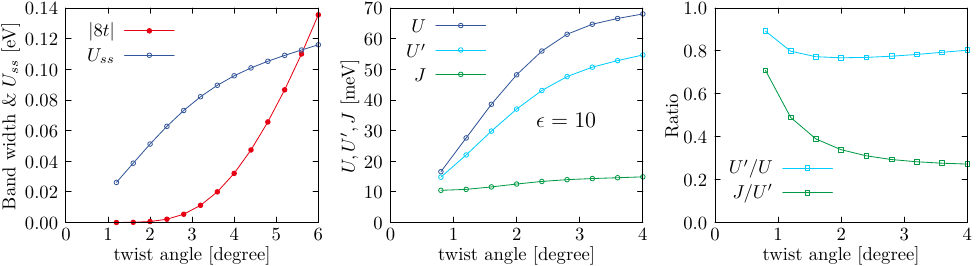}
\caption{Left: Onsite Coulomb interaction in the $s$-orbital $U_{ss}(0)$ as a function of twisting angle and comparison with bandwidth $W=|8t|$ (taking into account only the nearest neighbor hopping), plotted in eV. Right: For $p$-orbitals, strengths of intra-orbital Coulomb $U=U_{p_xp_x}(0)=U_{p_yp_y}(0)$, inter-orbital Coulomb $U'=U_{p_xp_y}(0)$ and Hund's coupling $J$, plotted in meV. }
\label{fig:Coulomb}
\end{figure}

The isolated nature of the lowest moiré bands (Fig. \ref{fig:corrugation_compare}) justifies a localized Wannier description. To resolve the arbitrary gauge phases of the numerical Bloch states, we project them onto localized Gaussian trial functions \cite{PhysRevB.56.12847}. This yields physically localized Wannier orbitals (Fig. \ref{fig:hopping_fit}, left) and their corresponding hopping parameters (Fig. \ref{fig:hopping_fit}, right). Integrating over these orbitals provides the Coulomb and exchange matrix elements, see for example refs. \cite{ToshiLuo,PhysRevB.75.224408}. Assuming a typical dielectric constant $\epsilon=10$ (which can be tuned via gate separation \cite{yang2025engineeringhubbardmodelsgated}), the onsite Coulomb interaction is of order $\sim 50$ meV, significantly exceeding the hopping terms. As shown in Fig. \ref{fig:Coulomb}, increasing $\theta$ increases the bandwidth until it crosses $U$, potentially driving a metal-insulator transition. Further details of the calculation, including the next-nearest neighbor Coulomb interactions which consistently show $2U_B<U_A+U_C$, are presented in {Appendix \ref{app:NN_Coulomb}}. 

\subsection{Symmetry breaking orders} 
To explore symmetry-breaking phases in the $p$-orbitals, we consider the moir\'e Hamiltonian:
\begin{align}
        & H_{p}=  - \sum_{i,\sigma,a}
        \big(t_{\sigma} c_{ip_a\sigma}^{\dagger}c_{i+\bm{a},p_a\sigma} +t_{\pi} c_{ip_a\sigma}^{\dagger}c_{i+\bar{\bm{a}},p_a\sigma}+\text{h.c.}\big)\nonumber \\
        & + \frac{1}{2}\sum_{i,a,\sigma}\big[U n_{i,p_a\sigma}n_{i,p_a\bar{\sigma}} -J c^\dagger_{ip_a\sigma}c_{ip_{\bar{a}}\bar{\sigma}}c^\dagger_{i,p_a\bar{\sigma}}c_{ip_{\bar{a}}\sigma} \nonumber \\
        & +\sum_{\sigma',a} \big( U'n_{ip_a\sigma}n_{ip_{\bar{a}}\sigma'}-J  c_{ip_a\sigma}^{\dagger}c_{ip_a\sigma'}c_{ip_{\bar{a}}\sigma'}^{\dagger} c_{ip_{\bar{a}}\sigma} \big)\big]     
\label{eq:2_orbital_Hubbard}
\end{align}
where $\bar{\uparrow}=\downarrow$, $\bar{\downarrow}=\uparrow$ for spins $\sigma$ and $p_{\bar{x}}=p_y$, $p_{\bar{y}}=p_x$ for orbitals. 
This model shares the same minimal two-orbital symmetry structure -- 
$C_{4z}$-related orbital doublets, anisotropic nearest-neighbor hopping, local multiorbital interactions -- of the $d_{xz}$, $d_{yz}$ models proposed for the pnictides \cite{PhysRevB.77.220503,PhysRevLett.101.237004,PhysRevB.79.134502},  which capture their key magnetic and pairing properties at half filling, though not necessarily their detailed Fermi-surface topology (see {Appendix \ref{app:pnictide}} for a careful comparison).

\begin{figure}[hbt]
\includegraphics[scale=0.55,trim={0 0.2cm 0 0 }, clip]{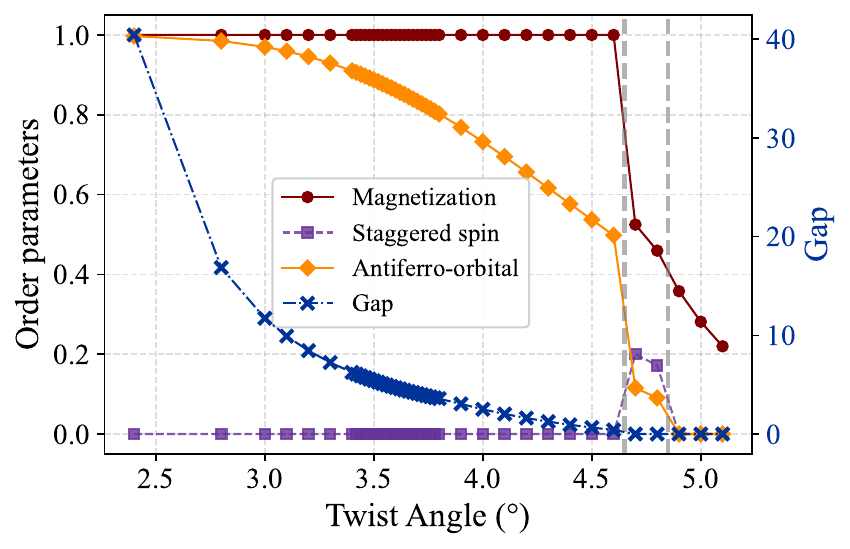}
\caption{At quarter filling of \eqref{eq:2_orbital_Hubbard} and in the parameter regime of $1<U/t_{\sigma}<100$, order parameters and gap are plotted as a function of twisting angle. The region with the densest data points correspond to the parameter regime where $U>U'> t_{\sigma}$ but all three are of the same order. When there are two sublattices $A, B$ per unit cell, magnetization is defined as $|\vec{S}_A+\vec{S}_B|/2$, staggered spin is defined as $|\vec{S}_A-\vec{S}_B|/2$, and antiferro-orbital order $\sum_{\sigma} (n_{Ap_x\sigma}+n_{Bp_y\sigma}-n_{Ap_y\sigma}-n_{Bp_x\sigma})$.}
\label{fig:HF}
\vspace{-0.5cm}
\end{figure}

Exploiting the tunability of vdW heterostructures via gating, we examine the quarter-filling physics of \eqref{eq:2_orbital_Hubbard} using self-consistent Hartree-Fock theory (numerical details described in {Appendix \ref{app:HF}}). Figure \ref{fig:HF} shows the resulting phase diagram with three distinct phases separated by first-order transitions. For $2.4^{\circ}\leq \theta\leq 4.6^{\circ}$, the system is an antiferro-orbital (AFO), ferromagnetic (FM) insulator with checkerboard ordering. 
In the strong-coupling regime, this AFO+FM state emerges intuitively because $U'<U$, such that orbital superexchange ($t^2/U'$) dominates over spin superexchange ($t^2/U$), favoring staggered orbitals, while Hund's coupling aligns spins. This orbital order is compatible with and complementary to the previous analysis where the onsite Coulomb interactions are irrelevant and only nearest neighbor Coulomb interaction was taken into account. 
The quasiparticle gap closes for $\theta>4.6^{\circ}$. Between $4.6^{\circ}<\theta<4.9^{\circ}$, kinetic-interaction competition stabilizes a stripe geometry with finite average and staggered magnetizations. For $\theta \geq 4.9^{\circ}$, the orbitals disorder, leaving a rapidly decaying, finite magnetization. It is expected that at even larger twisting angles, the system will become trivial symmetric metal.

\section{Discussions}
This work establishes $\Gamma$-valley square-lattice moiré systems as a versatile platform for simulating foundational models of high-$T_c$ superconductivity. These tunable van der Waals heterostructures provide a clean setting to benchmark competing orders and explore Hubbard model regimes inaccessible in bulk materials.
{The main results include: 
(i) The closed-form small-angle scalings for all parameters of the $s$- and $p$-orbital models of a generic $\Gamma$-valley square-lattice homobilayer (eqs.~\eqref{eq:t_s}--\eqref{eq:AFO}), verified against the first-principles parameters of twisted ZnF$_2$ in Figs.~\ref{fig:hopping_fit} and \ref{fig:Coulomb}. }
(ii) Because the consecutive moiré bands realize different effective models, a single device can potentially capture the physics of both cuprate and iron-pnictide families{, and these two regimes can be switched by moderate gating.} (iii) Focusing on the former, our candidate platform emulates cuprate physics by reaching the relevant interaction strength \cite{PhysRevB.79.235130} $U\sim 8t$ {at a twist angle $\theta\approx 5.6^{\circ}$, with $U/t$ increasing at smaller angles. (iv) At quarter filling of the $p$ doublet, the small-angle electrostatic analysis and Hartree-Fock theory consistently give an antiferro-orbital ferromagnetic insulator for $2.4^{\circ}\leq\theta\leq 4.6^{\circ}$. Antiferro-orbital fluctuations have been proposed as a pairing mechanism in two-orbital models of iron-based superconductors \cite{huang2017monolayer,PhysRevB.94.155127,PhysRevB.104.184507,JPSJ.85.114709}, but the ordered phase itself has not, to our knowledge, been observed; the present platform specifies a system and an angle window in which it should appear.} {Experimentally, this window corresponds to moir\'e periods $a_m\approx 9$--$5$ nm and, at three electrons per moir\'e cell (filled $s$ band plus one $p$ electron), to carrier densities $3/a_m^2\approx 3.5\times10^{12}$--$1.3\times10^{13}$ cm$^{-2}$, within reach of dual-gated devices. At this filling the phase would appear as an incompressible state  whose moir\'e unit cell is doubled, visible in scanning tunneling microscopy \cite{li2021imaging}.}

\emph{Note Added.} Subsequent to the initial posting of this work, Ref. \cite{shi2026gammavalleymoireplatformtunable} appeared independently, focusing on displacement-field-tuned $\Gamma$-valley moiré systems.

\vspace{0.3cm}

\begin{acknowledgments}
We are grateful to Junyeong Ahn, Gil Young Cho, Chaoxing Liu, Qimiao Si, Subir Sachdev, Zhida Song, Oskar Vafek, Cenke Xu, Hong Yao, and Jiabin Yu for enlightening conversations. Zhu-Xi thanks Jun-Kai Dong for discussions on Hartree-Fock calculations and Shijun Sun for related collaborations. This work was performed in part at the Aspen Center for Physics, which is supported by National Science Foundation grant PHY-2210452. T.K. was supported by JSPS KAKENHI Grant Number JP24K06968. A.V. acknowledges funding from NSF DMR-2220703 and from a Star-Friedman Challenge award at Harvard University.
\end{acknowledgments}

\bibliography{ref.bib}

\appendix

\begin{widetext}
\section{Mathieu functions and their asymptotic expansions}
\label{app:Matt}

We start from the effective 1D Hamiltonian of the variable-separated continuum Hamiltonian, 
\be
H_{\pm, x_i} = H_0 (-\i {\nabla}_i)\pm \frac{1}{2} T_0+(V_1\pm T_1) \cos (g_i x_i).
\ee
One can cast Schrödinger's equation in a standard form by reparametrizing $X_i=\frac{\pi}{a_m}x_i$, $A_{\pm}=\frac{m_*}{\hbar^2}\frac{a_m^2}{\pi^2}(2E_{\pm,i}\mp T_0)$ and $B_{\pm}=+\frac{m_* a_m^2}{\hbar^2\pi^2}(V_1\pm T_1)$. Then the Schrödinger's equation becomes Mathieu equation
\be
\frac{d^2}{dX_i^2}\psi_{\pm,i} (X_i) + \left[ A_{\pm} - 2B_{\pm} \cos (2X_i) 
\right] \psi_{\pm,i}(X_i) =0. 
\label{eq:Matt}
\ee
The full eigenergies are given by $E_\pm =E_{\pm,x}+ E_{\pm,y}$.  
Without loss of generality, we focus on the $-$ branch, omit the subscript in $A, B$ and assume $B>0$, but alternative choices can be analyzed using similar methods.  Solutions to \eqref{eq:Matt} are the cosine- and sine-elliptic Mathieu functions, reviewed in the table \ref{tab:Matt}.
\begin{table}[htbp]
\centering
\begin{tabular}{c|c|c|c}
Eigenvalue & Eigenfunction & Period & Parity \\
\hline
$a_{2n}(B)$ & $ce_{2n}(B,X_i)$ & $\pi$ & Even \\
\hline
$a_{2n+1}(B)$ & $ce_{2n+1}(B,X_i)$ & $2\pi$ & Even \\
\hline 
$b_{2n+1}(B)$ & $se_{2n+1}(B,X_i)$ & $2\pi$ & Odd \\
\hline
$b_{2n+2}(B)$ & $se_{2n+2}(B,X_i)$ & $\pi$ & Odd \\
\end{tabular}
\caption{Solutions to Mathieu equation.}
\label{tab:Matt}
\end{table} 
Without loss of generality we will focus on the $B>0$ branch, where the eigenvalues satisfy the relations $a_0<b_1<a_1<b_2<a_2<b_3\cdots$. These values are directly related to the energies at different high-symmetry points in the Brillouin zone.  

At the moir\'e $\Gamma_m$ point, the Bloch wavefunction is $\pi$-periodic, so we are only interested in the eigenvalues with even subscripts. The ground state will have energy $\varepsilon_s (\Gamma_m)=2a_0$, while the first and second excited states are degenerate with energy $\varepsilon_{p_x} (\Gamma_m)=\varepsilon_{p_y} (\Gamma_m)=a_0+b_2$, since we can have $a_0$ in the $X$ direction, $b_2$ in the $Y$ direction or vice versa.  As for the $M_m$ point at the moir\'e Brillouin zone corner, the Bloch wavefunction is $2\pi-$periodic. The ground state at this point has energy $\varepsilon_s (M_m)=2b_1$,  the first and second excited states have energy $\varepsilon_{p_x} (M_m)=\varepsilon_{p_y} (M_m)=a_1+b_1$ and is again twofold degenerate. At the $X_m$ ($Y_m$) point, the Bloch wavefunction is $\pi$-periodic in the $y$ ($x$)-direction, but $2\pi$ periodic in the $x$ ($y$)-direction. The lowest energy is $\varepsilon_s (X_m)=a_0+b_1$, and assuming without loss of generality $t_{\sigma}<t_{\pi}$, we further have excited state energies $\varepsilon_{p_x} (X_m)=\varepsilon_{p_y} (Y_m)=a_0+a_1$ and $\varepsilon_{p_y} (X_m)=\varepsilon_{p_x} (Y_m)=b_1+b_2.$ 
 In the asymptotic limit of $\theta \rightarrow 0$ or equivalently $B\rightarrow \infty$, physically intuitive analytical expressions can be obtained. 

In this limit, the eigenvalues of the Schrödinger's equation becomes \cite{olver2010nist} 
\be
a_m (B)\sim b_{m+1}(B) \sim -2 B +2(2m+1)\sqrt{B} + \mathcal{O}(1),
\ee
and 
\be
\begin{split}
b_{m+1}(B)-a_m(B) \sim & \frac{2^{4m+5}}{m!} \sqrt{\frac{2}{\pi}} B^{\frac{m}{2}+\frac{3}{4}} e^{-4\sqrt{B}} \\
& \qquad \times \left(1+ \mathcal{O}(B^{-1/2}) \right)
\end{split}
\ee
Taking $m=0$, this becomes 
\be
b_{1}(B)-a_0(B) \sim 32 \sqrt{\frac{2}{\pi}} B^{\frac{3}{4}} e^{-4\sqrt{B}} + \cdots
\ee
Taking $m=1$, this is
\be
b_{2}(B)-a_1(B) \sim 2^{9} \sqrt{\frac{2}{\pi}} B^{\frac{5}{4}} e^{-4\sqrt{B}} \left(1+ \mathcal{O}(B^{-1/2}) \right)
\ee
These formula lay the foundation for the hopping amplitude estimation. For example, $t_{\sigma}\propto[\varepsilon_{p_x}(X_m)-\varepsilon_{p_x}(\Gamma_m)]/4=(a_1-b_2)/4$, and 
$t_{\pi}\propto[\varepsilon_{p_x}(Y_m)-\varepsilon_{p_x}(\Gamma_m)]/4=(b_1-a_0)/4$. Using asymptotic expansion of the characteristic values, one can obtain
\be 
t_{\pi}\sim t,\quad t_{\sigma}\sim - 16B^{1/2}t_{\pi},
\ee
which is exactly equation (4) in the main text.

\section{Estimations of interactions}
\label{app:Gaussian}

This section details the evaluation of the Gaussian type integrals to estimate the interactions in the moiré Hubbard-type model,
\be
\frac{4\pi \epsilon}{e^2}U_{oo'}(\bm{d})=\int d\x  d\x' \frac{1}{2|\x-\x'|}\psi_o^2 (\x -\bm{d}) \psi_{o'}^2 (\x'). 
\label{eq:Coulomb_integral}
\ee
We start with the Coulomb interaction in $s-$orbital state,
\be
\psi_{s}^2 = \left(\frac{\alpha}{\pi}\right) e^{-\alpha (x^2+y^2)},\\
\ee
Change to center of mass and relative coordinates, $\bm{r}=\x-\x',$ $\bm{R}=\frac{1}{2}(\x+\x')$ such that 

\begin{equation}
\begin{split}
& \int d\x \int d\x' \frac{1}{2|\x-\x'|}\psi_s^2 (\x -\bm{d}) \psi_{s}^2 (\x') \\
= & \frac{\alpha^2}{2\pi^2} \int d\r \int d\R \frac{1}{|\r|} e^{-\alpha [(\R-\d-\r/2)^2+(\R+\r/2)^2]} \\
= & \frac{\alpha^2}{2\pi^2} \int_0^{\infty} dr \int_0^{2\pi} d\theta  e^{-\alpha (d^2+rd\cos\theta+r^2/2)}
\int_0^{\infty} R dR \int_0^{2\pi} d\phi e^{-\alpha (2R^2-2Rd\cos\phi)}\\
= & \frac{\alpha^2}{2\pi^2} e^{-\alpha d^2} \int_0^{\infty} dr\ 2\pi I_0(\alpha d r) e^{-\alpha r^2/2}
\int_0^{\infty} R dR\ 2\pi I_0 (2 \alpha d R) e^{-2\alpha R^2}\\
= & \frac{\alpha^2}{2\pi^2} e^{-\alpha d^2} \cdot 
\frac{\sqrt{2\pi}}{\sqrt{\alpha}} \pi e^{\alpha d^2/4} I_0(\alpha d^2/4) 
\cdot \frac{\pi}{2\alpha} e^{\alpha d^2/2}\\
= & \frac{\sqrt{\pi \alpha}}{2\sqrt{2}} I_0(\alpha d^2/4) e^{-\alpha d^2/4}
\end{split}
\end{equation}
where $I_0$ is the Bessel function of the first kind. So
\be
U_{ss}(\d)=\frac{e^2}{4\pi \epsilon}  \frac{\sqrt{\pi \alpha}}{2\sqrt{2}} I_0(\alpha d^2/4) e^{-\alpha d^2/4}
\ee
When $d\rightarrow 0$, we obtain the onsite contribution 
\be
U_{ss}(0)=\frac{e^2}{4\pi \epsilon}  \frac{\sqrt{\pi \alpha}}{2\sqrt{2}}.
\ee

Next we evaluate the intra-orbital Coulomb interaction for first excited states,
\be
\psi_{p_i}^2 =\left(\frac{2}{\pi}\right) \alpha^2  x_i^2 \ e^{-\alpha (x^2+y^2)}.
\ee
The relevant integral for intra-orbital Coulomb interaction is  
\begin{equation}
\begin{split}
& \int d\x \int d\x' \frac{1}{2|\x-\x'|}\psi_{p_x}^2 (\x -\bm{d}) \psi_{p_x}^2 (\x') \\
= & \frac{4\alpha^4}{2\pi^2} \int dx dx' x'^2 (x-d_x)^2  e^{-\alpha (x^2+d_x^2-2d_x x+x'^2)}\int dy dy' \frac{1}{\sqrt{(x-x')^2+(y-y')^2}} e^{-\alpha (y^2+d_y^2-2y d_y+y'^2)} \\
\end{split}
\end{equation}
Using the identity
\be
\frac{1}{\sqrt{(x-x')^2+(y-y')^2}} = \int \frac{d\bm{k}}{2\pi} \frac{e^{\i \bm{k}\cdot (\x-\x')}}{|\bm{k}|}
\label{eq:identity}
\ee
Above reduces to
\begin{equation}
\begin{split}
& \frac{4\alpha^4}{2\pi^2}\int \frac{d\bm{k}}{2\pi} \frac{1}{|\bm{k}|} \int dx dx' x'^2 (x-d_x)^2  e^{-\alpha (x^2+d_x^2-2d_x x+x'^2)+\i k_x(x-x')}\int dy dy'  e^{-\alpha (y^2+d_y^2-2y d_y+y'^2)+\i k_y(y-y')} \\
= & \frac{4\alpha^4}{2\pi^2} \frac{\pi}{\alpha} \int \frac{d\bm{k}}{2\pi} \frac{1}{|\bm{k}|} e^{\i k_y d_y - k_y^2/2\alpha}\int dx dx' x'^2 (x-d_x)^2  e^{-\alpha (x^2+d_x^2-2d_x x+x'^2)+\i k_x(x-x')} \\
= & \frac{4\alpha^4}{2\pi^2} \frac{\pi}{\alpha} \int \frac{d\bm{k}}{2\pi} \frac{1}{|\bm{k}|} e^{\i k_y d_y - k_y^2/2\alpha} \frac{\pi (k_x^2-2\alpha)^2}{16\alpha^5} e^{\i k_x d_x-k_x^2/2\alpha}\\
= &  \int \frac{d\bm{k}}{2\pi} \frac{1}{|\bm{k}|} e^{\i \bm{k}\cdot \bm{d} - \k^2/2\alpha} \frac{(k_x^2-2\alpha)^2}{8\alpha^2}  \\
\end{split}
\end{equation}
First assume $\d=(d,0)$, i.e. the separation is along the $x$-axis. Integral then becomes 
\begin{equation}
\begin{split}
U_A= &  \frac{e^2}{4\pi \epsilon} \int \frac{dk}{2\pi} d\theta e^{\i k d\cos\theta - k^2/2\alpha} \frac{(k^2\cos^2\theta-2\alpha)^2}{8\alpha^2}  \\
= & \frac{e^2}{4\pi \epsilon}  \frac{\sqrt{\pi\alpha}}{\sqrt{2}}e^{-\alpha d^2/4}\left[
\frac{7+\alpha^2d^4}{16}I_0 (\alpha d^2/4)-\frac{6+\alpha d^2 (5+\alpha d^2 (2+\alpha d^2)))}{16 \alpha d^2} I_1 (\alpha d^2/4)
\right]
\end{split}
\end{equation} 
Next assume $\d=(0,d)$. Integral then becomes
\begin{equation}
\begin{split}
U_C= &  \frac{e^2}{4\pi\epsilon} \int \frac{dk}{2\pi} d\theta e^{\i k d\sin\theta - k^2/2\alpha} \frac{(k^2\cos^2\theta-2\alpha)^2}{8\alpha^2}  \\
= &\frac{e^2}{4\pi\epsilon}  \frac{\sqrt{\pi\alpha}}{\sqrt{2}}e^{-\alpha d^2/4}
\left[
\frac{7}{16}I_0 (\alpha d^2/4)+\frac{\alpha d^2 -6}{16 \alpha d^2}  I_1 (\alpha d^2/4)
\right]. 
\end{split}
\end{equation}
It is also possible to evaluate the integral for arbitrary $\d=(d_x,d_y)$ but the result is tedious and not physically intuitive, so we will omit it here.
In the onsite $d\rightarrow 0$ limit, the Coulomb interaction becomes
\be
U_{p_xp_x}(0)=\frac{e^2}{4\pi \epsilon}  \frac{\sqrt{\pi\alpha}}{\sqrt{2}}
\left[
\frac{7}{16}-\frac{6}{128}  \right] = \frac{25}{32}\frac{e^2}{4\pi \epsilon}  \frac{\sqrt{\pi\alpha}}{2\sqrt{2}}.
\ee

Next we turn to the inter-orbital interaction $U_{p_xp_y}$.  Using again the identity \eqref{eq:identity},
\begin{equation}
\begin{split}
 & \int d\x \int d\x' \frac{1}{2|\x-\x'|}\psi_{p_x}^2 (\x -\bm{d}) \psi_{p_y}^2 (\x') \\
= & \frac{4\alpha^4}{2\pi^2} \int \frac{d\bm{k}}{2\pi} \frac{1}{|\bm{k}|} \int dx dx'  (x-d_x)^2  e^{-\alpha (x^2+d_x^2-2d_x x+x'^2)+\i k_x(x-x')}\int dy dy'  y'^2 e^{-\alpha (y^2+d_y^2-2y d_y+y'^2)+\i k_y(y-y')}\\
= & \frac{4\alpha^4}{2\pi^2} \int \frac{d\bm{k}}{2\pi} \frac{1}{|\bm{k}|} \cdot \frac{\pi (k_y^2-2\alpha)}{4\alpha^3} \frac{\pi (k_x^2-2\alpha)}{4\alpha^3} e^{\i \bm{k}\cdot \bm{d}-\bm{k}^2/2\alpha}
\end{split}
\end{equation}
Assume $\d=(d,0)$, i.e. the separation is along the $x$-axis. Integral then becomes 
\begin{equation}
\begin{split}
U_B= & \frac{e^2}{4\pi\epsilon} \int \frac{dk}{2\pi} d\theta e^{\i k d\cos\theta - k^2/2\alpha} \frac{(k^2\cos^2\theta-2\alpha)(k^2\sin^2\theta-2\alpha)}{8\alpha^2}  \\
= & \frac{e^2}{4\pi\epsilon} \frac{\sqrt{\pi\alpha}}{\sqrt{2}} e^{-\alpha d^2/4} \left[
\frac{(4+\alpha d^2)}{16} I_0 (\alpha d^2/4) + \frac{6+\alpha d^2 (2-\alpha d^2)}{16 \alpha d^2} I_1 (\alpha d^2/4)\right]\\
\end{split}
\end{equation}
The onsite contribution is straightforward to find 
\be
U_{p_xp_y}(0) = \frac{19}{32}\frac{e^2}{4\pi\epsilon} \frac{\sqrt{\pi\alpha}}{2\sqrt{2}} 
\ee
To find the orbital ordering in the asymptotic limit, we need to compute
\be
\begin{split}
& U_A+U_C-2U_B \\
= & \frac{e^2}{4\pi\epsilon} \frac{\sqrt{\pi\alpha}}{\sqrt{2}} e^{-\alpha d^2/4} \bigg[
I_0(\alpha d^2/4) \left(\frac{7+\alpha^2d^4}{16}  +\frac{7}{16}-2\frac{(4+\alpha d^2)}{16} \right)\\
& \quad \quad + I_1(\alpha d^2/4)\left(-\frac{6+\alpha d^2 (5+\alpha d^2 (2+\alpha d^2)))}{16 \alpha d^2} +\frac{\alpha d^2 -6}{16 \alpha d^2}-2 \frac{6+\alpha d^2 (2-\alpha d^2)}{16\alpha d^2} \right)
\bigg] \\
= & \frac{e^2}{4\pi\epsilon} \frac{\sqrt{\pi\alpha}}{\sqrt{2}} e^{-\alpha d^2/4} \frac{1}{16} \bigg[
\left(6-2\alpha d^2+\alpha^2 d^4 \right) I_0(\alpha d^2/4) +\frac{1}{\alpha d^2} \left(-24-8\alpha d^2-\alpha^3 d^6\right)  I_1(\alpha d^2/4) 
\bigg] \\
\end{split}
\ee
Perform the asymptotic expansion up to order $n_*=5,$
\be
I_{\nu} (x) \sim  \frac{e^{x}}{\sqrt{2\pi x}} \sum_{n=0}^{n_*} (-1)^n \frac{\prod_{m=1}^{2n+1} (4\nu^2-m^2)  }{n! (8x)^n},
\ee
the result simplifies to, at leading order,
\be
U_A+U_C-2U_B = \frac{e^2}{4\pi\epsilon} \frac{57}{8\alpha^2 d^5} + O(\alpha^{-3}d^{-7}),
\ee
which is always positive.

We further analyze the hybridization between the $s$- and the $p$-orbitals. The relevant integral is 
\begin{equation}
\begin{split}
U_{sp_x}(\d) = & \frac{e^2}{4\pi\epsilon} \int d\x \int d\x' \frac{1}{2|\x-\x'|}\psi_{p_x}^2 (\x -\bm{d}) \psi_{s}^2 (\x') \\
= & \frac{e^2}{4\pi\epsilon} \frac{\alpha^3}{\pi^2} \int dx dx' x'^2   e^{-\alpha (x^2+d_x^2-2d_x x+x'^2)}\int dy dy' \frac{1}{\sqrt{(x-x')^2+(y-y')^2}} e^{-\alpha (y^2+d_y^2-2y d_y+y'^2)} \\
= & \frac{e^2}{4\pi\epsilon} \frac{\alpha^3}{\pi^2} \int \frac{d\bm{k}}{2\pi}\frac{1}{|\bm{k}|} \int dx dx' x'^2   e^{\i k_x  (x-x')}  e^{-\alpha (x^2+d_x^2-2d_x x+x'^2)}\int dy dy' e^{\i k_y  (y-y')} e^{-\alpha (y^2+d_y^2-2y d_y+y'^2)} \\
= & \frac{e^2}{4\pi\epsilon} \frac{\alpha^3}{\pi^2} \int \frac{d\bm{k}}{2\pi}\frac{1}{|\bm{k}|} 
\frac{\pi (2\alpha-k_x^2)}{4\alpha^3} e^{\i d_x k_x -k_x^2/2\alpha}
e^{\i d_y k_y -k_y^2/2\alpha} \frac{\pi}{\alpha}. \\
\end{split}
\end{equation} 
In particular the onsite term is
\be
U_{sp_x}(0)= \frac{e^2}{4\pi\epsilon} \frac{\alpha^3}{\pi^2} \frac{3\pi^{5/2}}{8\sqrt{2}\alpha^{5/2}} =\frac{3}{4} \frac{e^2}{4\pi\epsilon}\frac{\sqrt{\pi\alpha}}{2\sqrt{2}},
\ee
which satisfies $U_{ss}(0)>U_{p_xp_x}(0)>U_{sp_x}(0)>U_{p_xp_y}(0)$ as expected.

Finally we calculate the other quartic interactions that are not of Coulomb type and will only focus on the onsite contributions. The Hund's coupling is, 
\be
\frac{4\pi\epsilon}{e^2} J_H(0) = \int d\x \int d\x' \frac{1}{|\x-\x'|} \psi_{o}^*(x)\psi_{o'}(x)\psi^*_{o'}(x')\psi_{o}(x')
\ee
The pair hopping term is
\be
\frac{4\pi\epsilon}{e^2} J'(0) = \int d\x \int d\x' \frac{1}{|\x-\x'|} \psi_{o}^*(x)\psi_{o'}(x)\psi^*_{o}(x')\psi_{o'}(x')
\ee
Since all the wavefunctions are real, we have $J_H=J'\equiv J$. Taking $o=p_x, o'=p_y$, the integral is 
\be
\begin{split}
& \int d\x \int d\x' \frac{1}{|\x-\x'|} \psi_{o}^*(x)\psi_{o'}(x)\psi^*_{o}(x')\psi_{o'}(x') \\
= & \frac{4\alpha^4}{\pi^2} \int d\x \int d\x' \frac{1}{|\x-\x'|} xx'yy'e^{-\alpha (x^2+y^2+x'^2+y'^2)}\\
= & \frac{4\alpha^4}{\pi^2}\int \frac{d\k}{2\pi} \frac{1}{|\bm{k}|} \int d\x \int d\x'  xx'yy' e^{\i k_x (x-x')+\i k_y (y-y')}e^{-\alpha (x^2+y^2+x'^2+y'^2)}\\
= & \frac{4\alpha^4}{\pi^2} \frac{\pi^2}{16\alpha^6}\int \frac{d\k}{2\pi} \frac{1}{|\bm{k}|}  k_x^2 k_y^2 e^{-(k_x^2+k_y^2)/2\alpha}\\
= & \frac{4\alpha^4}{\pi^2} \frac{\pi^2}{16\alpha^6}  \frac{1}{2\pi}  \frac{3\alpha^{5/2}\pi^{3/2}}{4\sqrt{2}} = \frac{3}{16} \frac{\sqrt{\pi\alpha}}{2\sqrt{2}}.
\end{split}
\ee
Therefore $J_{p_xp_y}(0)=\frac{3}{16} \frac{e^2}{4\pi\epsilon} \frac{\sqrt{\pi\alpha}}{2\sqrt{2}}< U_{p_xp_y}(0)$.

\section{First-principle calculation details}
\label{app:DFT}

This section explains in more detail the setup for first-principle calculations for ZnF$_2$. We performed density function theory calculations using the Quantum Espresso package \cite{Giannozzi_2009,Giannozzi_2017} and implemented the VDW-DF2-B86R choice of the exchange-correlation functional for van der Waals forces \cite{PhysRevB.89.121103}. Pseudopotentials are of type PBE and from PS library \cite{DALCORSO2014337, PSL}. A thick supercell in the $z$-direction of $100$ Angstrom is used in order to handle the two-dimensional material. 
The monolayer band structure and orbital information can be found in figs. \ref{fig:mono_band} and \ref{fig:orbital}.
\begin{figure}[htbp]
\centering
\includegraphics[scale=0.6]{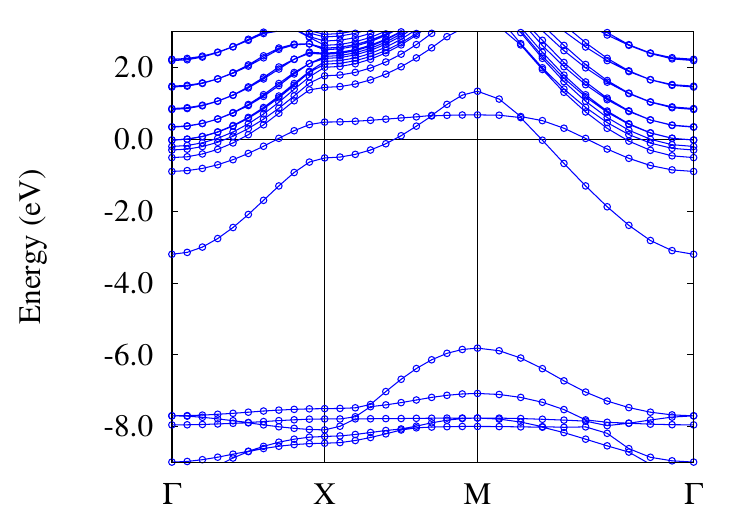}
\caption{Monolayer band structure showing conduction band minimum (CBM) at the $\Gamma$ point and valence band maximum (VBM) at the $M$ point. }
\label{fig:mono_band}
\end{figure}

\begin{figure}[htbp]
\centering
\includegraphics[scale=0.2]{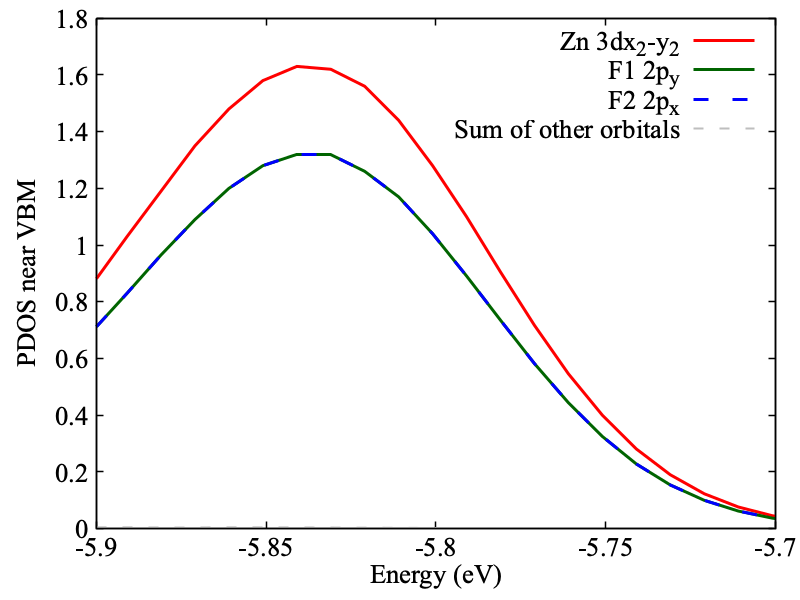}
\includegraphics[scale=0.2]{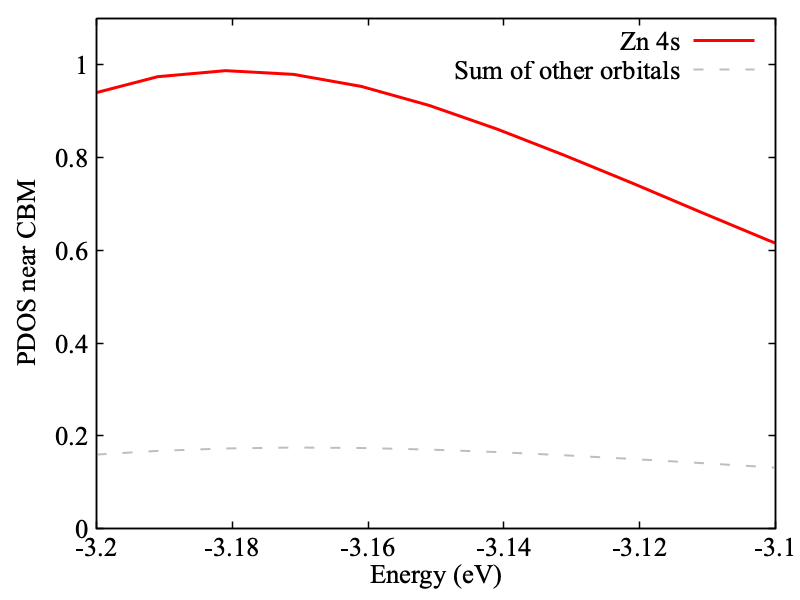}
\caption{Left: Projected density of states at the VBM - the dominating orbitals are Zinc $3d_{x^2-y^2}$ and Fluoride $2p_x$, $2p_y$. Right: Projected density of states at CBM, dominating orbital is Zinc $1s$. The summation of contributions from all other orbitals in all atoms is significantly smaller than Zn $4s$.}
\label{fig:orbital}
\end{figure}

For bilayer, we incorporate the corrugation effects between planes, while assuming a rigid material plane. The optimal interlayer distance modulates in the following symmetry-constrained way: 
\be
d(\x) = d_0 +\sum_n d_n \left [
\sum_{c=0}^3 \cos \left(2\pi \bm{x}\cdot (C_{4z}^c\cdot \bm{v}_n) \right)
\right], 
\label{eq:dz}
\ee
where $n\in\mathbb{Z}$ and $d_n$ are undetermined coefficients for fitting. $\bm{v}_n$ is the $n$-th nearest neighbor vector which has the smallest positive angle with respect to the $x$-axis: $\bm{v}_1=(1,0)$, $\bm{v}_2=(1,1)$, $\bm{v}_3=(2,0)$, etc. And $C_{4z}^c\cdot \bm{v}_n$ is the action of fourfold rotation on vector $\bm{v}_n$ by $c$-times. For each position $\x$, the optimal $d_z$ is found by minimizing the total bilayer energy - the latter first decrease and then increase with respect to the interlayer distance. The parameters in equation \eqref{eq:dz} are fitted to be {(in units of $10^{-3}a_0$)}:
\be
\begin{split}
& d_0=723.75,\quad d_1=36.9956,\quad d_2=38.7193,\\
& d_3=4.91071, \quad d_4= 2.13811.
\end{split}
\ee
We plot the spatial profile of $d_z$ in fig. \ref{fig:dz}.
\begin{figure}[htbp]
\centering
\includegraphics[scale=0.5]{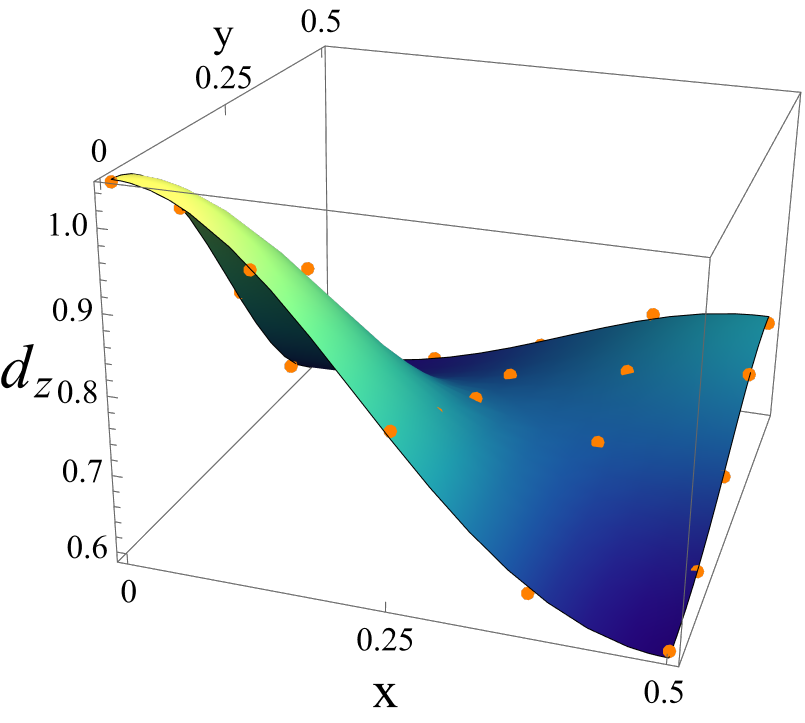}
\caption{Optimal interlayer distance as a function of relative shift between layers in the first quadrant.}
\label{fig:dz}
\end{figure}
The symmetry-allowed intralayer and interlayer moir\'e potentials, defined in (1) of the main text, can also be expanded in  harmonics, 
\be
\begin{split}
& V(\x) = V_0 +\sum_{n=1} V_n \left [
\sum_{c=0}^3 \cos \left(2\pi \bm{x}\cdot (C_{4z}^c\cdot \bm{v}_n) \right)
\right] \\
& T(\x) = T_0 +\sum_{n=1} T_n \left [
\sum_{c=0}^3 \cos \left(2\pi \bm{x}\cdot (C_{4z}^c\cdot \bm{v}_n) \right)
\right],
\end{split}
\label{eq:parameters}
\ee
For example, when $n=1,$ we just get back the terms in (2) of the main text. The numbers for $V_n$ and $T_n$'s are presented in table \ref{tab:fitting}, and the resultant moiré potential is in Fig. 3 of the main text. 
\begin{table}[htbp]
\centering
\begin{tabular}{c|c|c|c|c|c}
 & $n=1$ & $n=2$ & $n=3$ & $n=4$ & $n=5$ \\
\hline
$V_n$ & $-40.17013$ & $-55.305$ & $10.8649$ & $15.7927$ & $4.58234$ \\
\hline
$T_n$ &  $-3.42321$ & $-23.6122$ &  $-17.0831$ &  $-14.7755$ &  $-1.42366$ 
\end{tabular}
\caption{Fitted strengths of the moiré potentials in the twisted bilayer Hamiltonian for the $n$-th harmonics, in units of meV. As for the constant terms, $V_0=2.87056$ eV and $T_0=0.628548$ eV. The fact that $|T_1|<|T_2|$ originates from the modulation of $d_z$ - the rigid case leads to $|T_1|>|T_2|$ instead.}
\label{tab:fitting}
\end{table}
If instead $d_z$ was held to a constant value independent of $\bm{x}$, interlayer tunneling is expected to be large at trivial stacking, where the distance between the sites in the two layers is the smallest. In fig. \ref{fig:no_corrugation}, we further plot the moiré potential calculated from a constant $d_z$ chosen to be the average of the modulating $d_z(\bm{x})$ in moiré unit cell). 
\begin{figure}[htbp]
\centering
\includegraphics[scale=0.4]{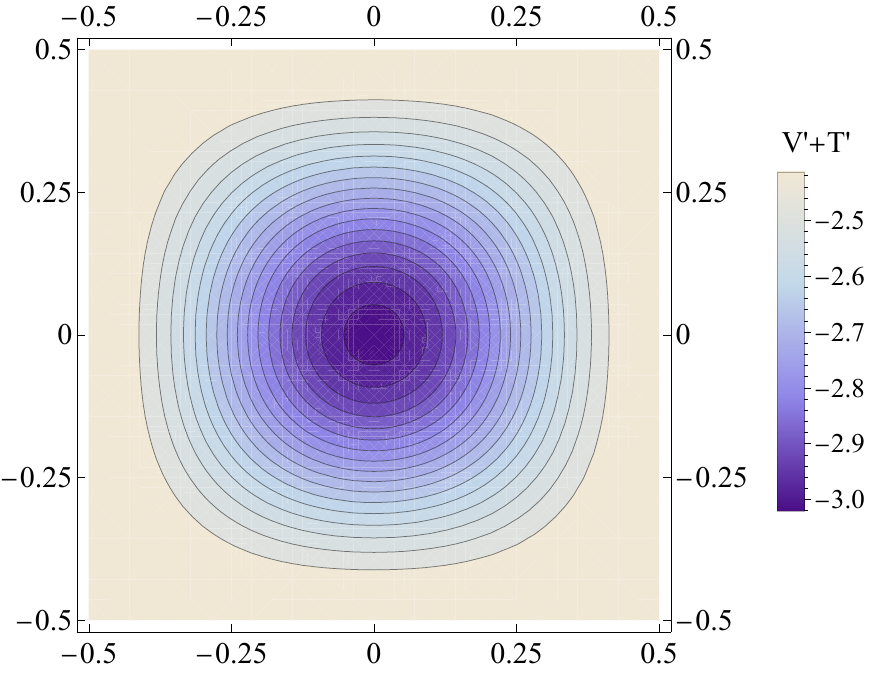}
\caption{
Real-space profiles for moiré potential of twisted bilayer ZnF$_2$, plotted as a function of normalized coordinates within the moiré unit cell in eV. In contrast to Fig. 3 in the main text, here the interlayer distance $d_z$ is taken to be uniform. The $d_z$ modulation splits the moiré potential minimum from one point at the unit cell center. 
}
\label{fig:no_corrugation}
\end{figure}

This effective modeling of the moiré system has been estimated using approximations of untwisted bilayers with varying relative shifts between the two layers, which requires fittings of the band structures. Novel alternative methods which do not require fitting are available have recently been proposed, which can provide complementary perspectives \cite{PhysRevB.107.075123, zhang2024universalmoiremodelbuildingmethodfitting,liu2025symmetryenforcedmoiretopology}. We will leave this for future investigations. 

\section{Band structures at larger twisting angles}
\label{app:higher_angle}

In figures 3 and 4 of the main text, we focused on the small-angle band structures of the continuum Hamiltonian. Such regime is chosen to align with the asymptotic analysis of the Mathieu functions and the estimation of kinetic and interaction energy scales. At larger twisting angles, higher harmonics become more important in the continuum Hamiltonian, which in turn lead to stronger longer-ranged hoppings in the effective model on the superlattice, see for example fig. 5 of the main text. Fig.\ref{fig:higher_deg_bands} shows additional band structure plots at larger twisting angles. 
\begin{figure}[htbp]
\centering
\includegraphics[scale=0.295]{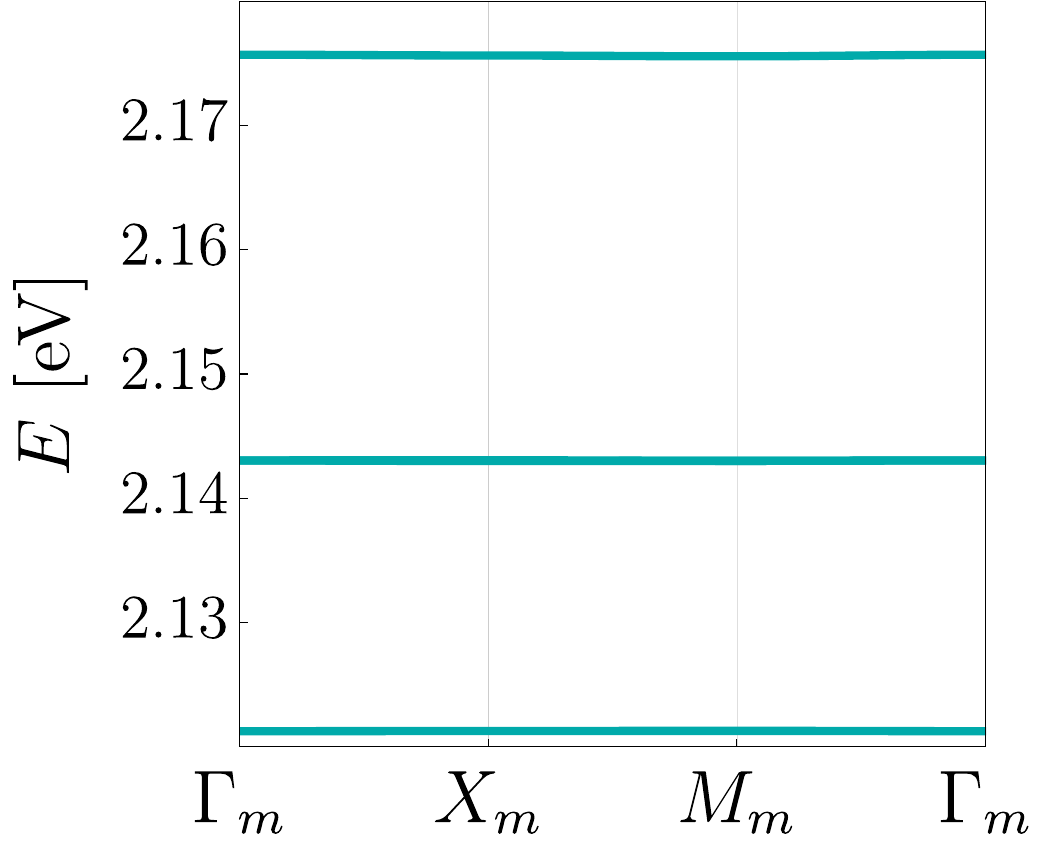}
\quad
\includegraphics[scale=0.3]{fig_s_orbital.pdf}
\quad 
\includegraphics[scale=0.3]{fig_p_orbital.pdf}\\
\includegraphics[scale=0.3]{fig_bands_2.pdf}
\quad
\includegraphics[scale=0.3]{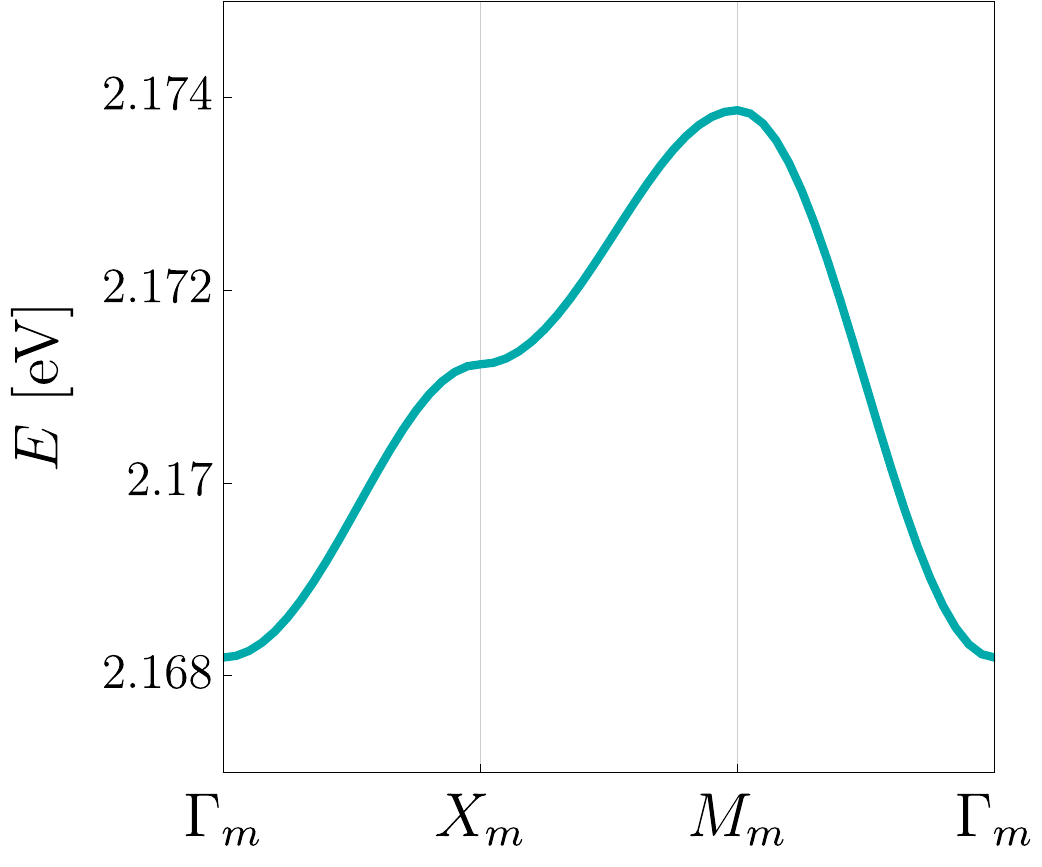}
\quad 
\includegraphics[scale=0.3]{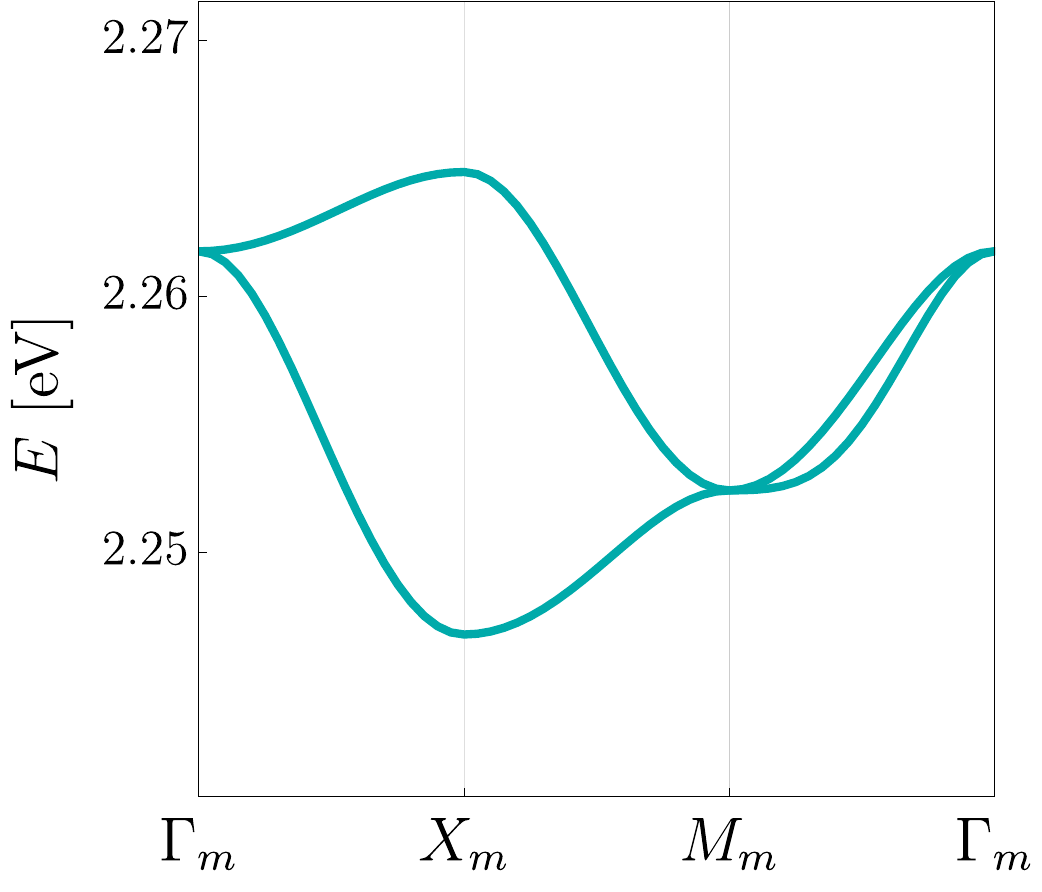}\\
\caption{Left column: Lowest bands of the continuum Hamiltonian at larger twisting angles. Middle column: zoomed-in view for the lowest s-orbital moiré band. Right column: the next lowest, $p_x$, $p_y$ moiré bands.  Top row: $\theta=1^{\circ}$. Bottom row: $\theta=2^{\circ}$.}
\label{fig:higher_deg_bands}
\end{figure}
Upon increasing twisting angles, the lowest moiré band remains qualitatively the same except for the increase in bandwidth. For the $p_x, p_y$ bands, however, increase of twisting angle further induces small but visible splittings along the $\Gamma_m-M_m$ line. This is due to the symmetry-allowed inter-orbital hoppings between next nearest neighbors; in contrast, nearest-neighbor inter-orbital hoppings are forbidden by reflection symmetries.

\section{Coulomb interactions in ZnF$_2$}
\label{app:NN_Coulomb}
The quartic terms in the effective moiré Hubbard model of twisted bilayer ZnF$_2$ can be estimated from the Wannier functions. We assume the interactions are screened by top and bottom metallic gates, separated from the sample with distance $\xi$, such that the Coulomb interaction is 
\be
\begin{split}
 U_{oo'} (\bm{d})= & \int d^2\x d^2\x' |\psi_o (\x-\bm{d})|^2 V(\bm{x}-\bm{x}') |\psi_{o'} (\x')|^2\\
= & \int \frac{d^2 \bm{q}}{(2\pi)^2} \rho_o (\bm{q}) \rho_{o'} (-\bm{q}) \frac{e^2}{2\epsilon \epsilon_0 q} \tanh (q \xi) e^{-\mathrm{i}\bm{q}\cdot \bm{d}},
\end{split}
\ee
where $V(\bm{x}-\bm{x}')=\frac{e^2}{\epsilon_0\epsilon |\bm{x}-\bm{x}'|}$ and $\epsilon$ is taken to be $10$ throughout. 
The orbitals $o, o'$ can be $s, p_x, p_y$. 
Similarly, the onsite Hund's coupling $J$ for the $p$-orbital is
\be
J_{p_x,p_y}=\int d^2\x d^2\x' \psi_{p_x}^*(\x)\psi_{p_y}(\x)   V(\bm{x}-\bm{x}') \psi_{p_y}^* (\x')\psi_{p_x}(\x').
\ee 
In Fig. \ref{fig:Unn_ABC}, we plot the nearest neighbor Coulomb interactions $U_A=U_{p_xp_x}(a_m,0),$ $U_B=U_{p_xp_y}(a_m,0),$ $U_C=U_{p_xp_x}(0,a_m)$ as a function of twisting angle. 
\begin{figure}[htbp]
\centering
\includegraphics[scale=1]{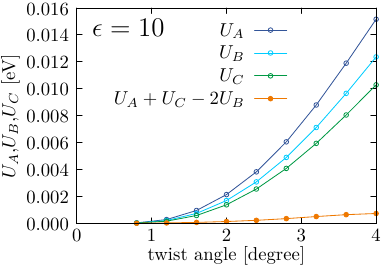}
\caption{$U_A+U_C-2U_B>0$ indicates antiferro-orbital order, consistent with the asymptotic analysis.} 
\label{fig:Unn_ABC}
\end{figure}

In the numerical calculations,the lowest $s$-orbital moir\'e band and two $p_x, p_y$-bands that follow are treated as separate, i.e. potential hybridization has been ignored. 
At $\theta=1^{\circ}$, the moiré band gap between the $s$ and $p$ bands is of order 25 meV, and the hybridization due to Coulomb interaction is expected to live below the $U_{ss}$ value of 15 meV (see for example figures 3 and 6 in the main text), so there doesn't seem to be hybridization issue. However, more careful analysis will need to be carried out to give a complete answer to this question, which we leave for future work.

\section{Comparison with the minimal pnictide Fermi surface}
\label{app:pnictide}

In this section, we elaborate on the relationship between our $p_x$, $p_y$-orbital bands and the minimal $d_{xz},d_{yz}$ model in Ref.~\cite{PhysRevB.77.220503} from a Fermiology perspective. In fig. \ref{fig:Fermiology}, we compare the half-filling Fermi surfaces of (left) the non-interacting tight-binding model of twisted bilayer ZnF$_2$, with hopping parameters extracted from fig. 5 in the main text, and (right) the iron pnictides. 
\begin{figure}[htbp]
\centering
\includegraphics[scale=0.5]{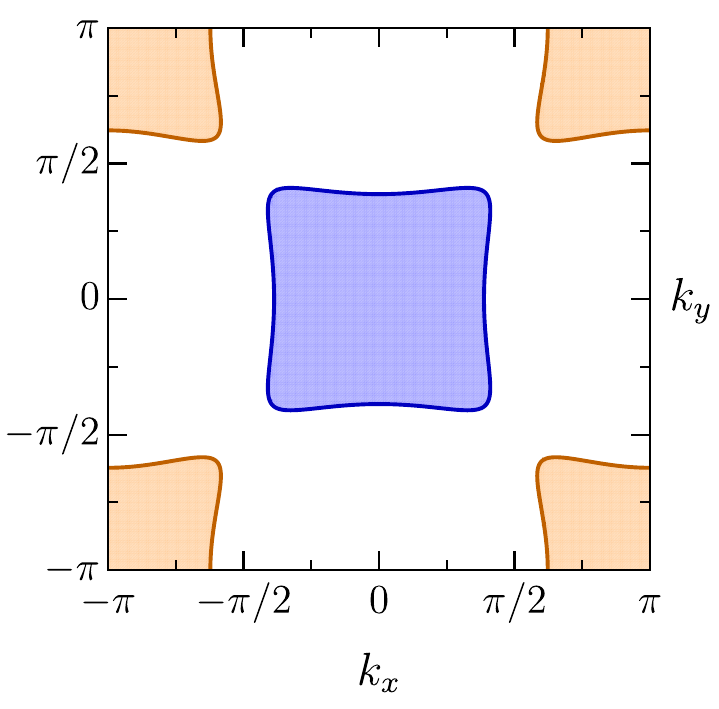}
\qquad
\includegraphics[scale=0.5]{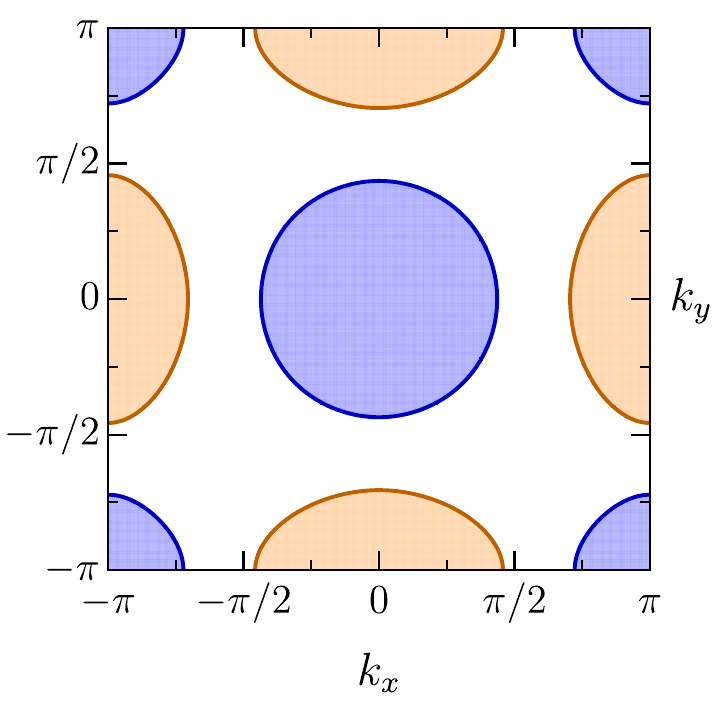}
\caption{Left: Electron and hole pockets for twisted bilayer ZnF$_2$ at $\theta=3.6^{\circ}$.  The hopping parameters have the ratios $t_{\pi}/t_{\sigma}=-0.3597$, $t_{nnn}^{xx}/t_{\sigma}=0.0468$ and $t_{nnn}^{xy}/t_{\sigma}=-0.1143$. Right: By comparison, iron pnictides are expected to have the pockets located as in the figure. The figure can be produced using the same functional form of the $p_x, p_y$-tight binding Hamiltonian, but with different choices of parameters:  $t_{\pi}/t_{\sigma}=-0.3$, $t_{nnn}^{xx}/t_{\sigma}=0.8$ and $t_{nnn}^{xy}/t_{\sigma}=-0.5$.}
\label{fig:Fermiology}
\end{figure}

At half filling, the fitted parameters of moiré ZnF$_2$ typically give a compensated semimetal with a single hole pocket at $\Gamma_m$ and a single electron pocket at $M_m$. This differs from the pnictide
Fermi surface, which in the same Brillouin zone consists of four pockets---hole pockets at $\Gamma$ and $M$ and electron pockets at $(\pi,0)$ and
$(0,\pi)$---connected by the $(\pi,0)$ nesting that is believed to underlie the stripe
spin-density-wave and $s_\pm$ pairing scenarios. Retaining the moir\'e
nearest-neighbor anisotropy but enlarging the next-nearest-neighbor hoppings can recover this arrangement. In twisted ZnF$_2$, within the parameter regime we examined, the NNN hoppings are not large enough to reach the pnictide physics. We note, however, that in the moir\'e setting, the relative strength between NNN and NN hoppings is in principle continuously tunable: shallower moiré potentials and larger twisting angles enhance it. Whether the threshold can be crossed in some specific $\Gamma$-valley
materials is left for future
investigation.

\section{Hartree-Fock details}
\label{app:HF}

The relevant interaction terms for the $p$-orbitals in equation (7) can be re-organized as 
\be
\begin{split}
H_{\text{onsite}}=\frac{1}{2}\sum_{a,\sigma}\bigg[U n_{a\sigma}n_{a\bar{\sigma}} 
-J c^\dagger_{a\sigma}c_{\bar{a}\bar{\sigma}}c^\dagger_{a\bar{\sigma}}c_{\bar{a}\sigma}\\
+\sum_{\sigma'} \left( U'n_{a\sigma}n_{\bar{a}\sigma'}
-J  c_{a\sigma}^{\dagger}c_{a\sigma'}c_{\bar{a}\sigma'}^{\dagger} c_{\bar{a}\sigma} \right)
\bigg],
\end{split}
\ee
The mean field tensors are defined as $\rho_{a\sigma,a'\sigma'}=\langle c_{a\sigma}^{\dagger} c_{a'\sigma'}\rangle.$ The range of twisting angles we focus on is $2.4^{\circ}\leq \theta \leq 5.1^{\circ}$, such that $1<U/t_{\sigma}<100$. We reset the hopping amplitude $t_{\sigma}=1$, and all the other parameters $t_{\pi}, U, U', J$ are renormalized to be $t_{\pi}/t_{\sigma}, U/t_{\sigma}, U'/t_{\sigma}, J/t_{\sigma}.$ The system size is chosen to be $140\times 140.$ To ensure a robust Hartree-Fock procedure to find the global instead of local minimum, we incorporate the following strategy for the consistent mean field theory solution:
\begin{itemize}
\item[(i)] For a variety of idealized initial ansätze with different symmetry breaking patterns (which include Néel order, ferromagnetic order, ferro-orbital and antiferro-orbital order, spin/orbital stripe order, charge density waves, and coexistence of those), as well as 20 random ansätze, first run the standard self-consistent calculation for each parameter combinations to obtain an initial set $\mathcal{S}_1$ of converged low-energy solutions.

\item[(ii)] In each run, tolerance of $10^{-5}$ is set for convergence. We exploit a temperature annealing schedule to avoid getting trapped in the first local minimum the algorithm finds. The inverse temperature $\beta$ is initially set to $5$, then at iteration 25, $\beta$ is increased to $15$, and then at iteration 75, $\beta$ is increased to $50$. We also utilize the Direct Inversion in the Iterative Subspace (DIIS) convergence acceleration technique \cite{PULAY1980393,pulay1982improved}, which incorporates a history of previous solutions and the corresponding errors, and choose the new solution to be the best linear combination of previous solutions which makes the corresponding error as close to zero as possible. This is to avoid oscillations and slow convergence that can arise from simple linear mixings between input and output solutions.  
\item[(iii)] For parameters $\{t_y(\theta), U(\theta), U'(\theta), J(\theta)\}$, use the  solutions in $\mathcal{S}_1$ from the neighboring twisting angles $\theta-\Delta \theta$ and $\theta+\Delta \theta$ as the initial guesses for the new calculation with angle $\theta$. This refinement run is to prevent the case where the idealized initial ansätze for certain phases do not lead to converged solutions in that phase, while an imperfect solution in that phase actually exists. This step will lead to an additional set of solutions $\mathcal{S}_2$.

\item[(iv)] For each previously found solution in sets $\mathcal{S}_1$, $\mathcal{S}_2$, we calculate the order parameters and assign it a label (e.g., ``Spin-AFM'', ``CDW'', ``Orbital-Ferro''). For each label, we store the single solution  with the highest order parameter magnitude it has found anywhere on the phase diagram. Then for all twisting angles, we use these global champions in each phase as the initial ansätze. This gives yet another additional set of solutions $\mathcal{S}_3.$ This step is added because there might be phases isolated in a corner of the phase diagram and the previous refinement runs may not be able to capture such candidates at the opposite corners.
\end{itemize}
The sets $\mathcal{S}_1, \mathcal{S}_2, \mathcal{S}_3$ will then be merged to find the lowest energy state. 

Fig. \ref{fig:HF_total_energy} plots the energies as a function of twisting angle of low-energy solutions with different unit cell types. 
\begin{figure}[htbp]
\includegraphics[scale=0.6]{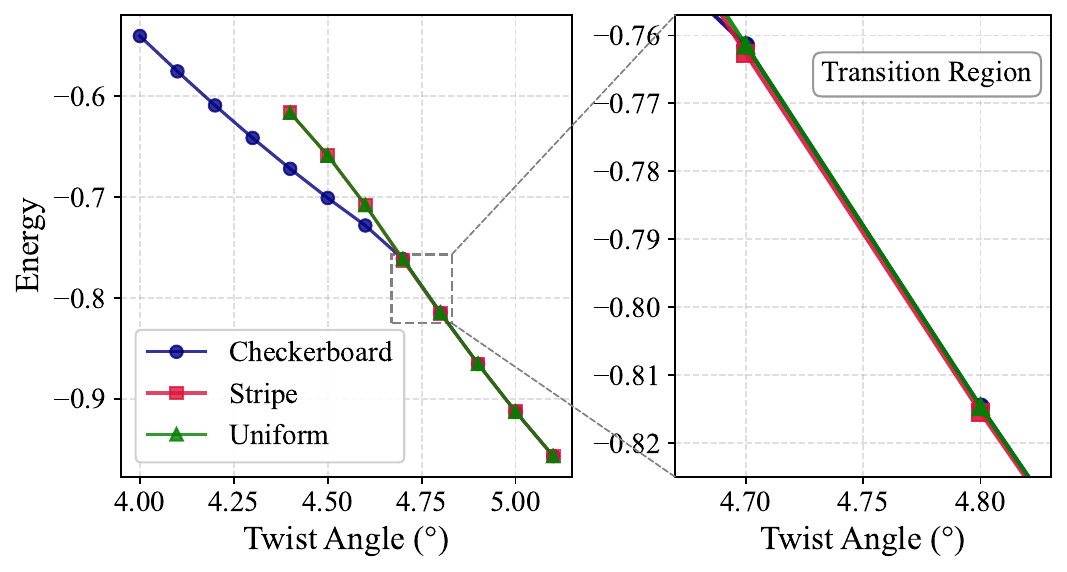}
\caption{Total energy as a function of twisting angle near the transition regions. The separation between the stripe geometry and other geometries is of order $10^{-3}$, while the error bar is of order $10^{-5}$.}
\label{fig:HF_total_energy}
\end{figure}
Figures \ref{fig:HF_band_structure_1}, \ref{fig:HF_band_structure_2}, \ref{fig:HF_band_structure_3} and \ref{fig:HF_band_structure_4} plot flavor-resolved band structures in various phases, in the folded Brillouin zone. 
\begin{figure}[htbp]
\centering
\includegraphics[scale=0.65]{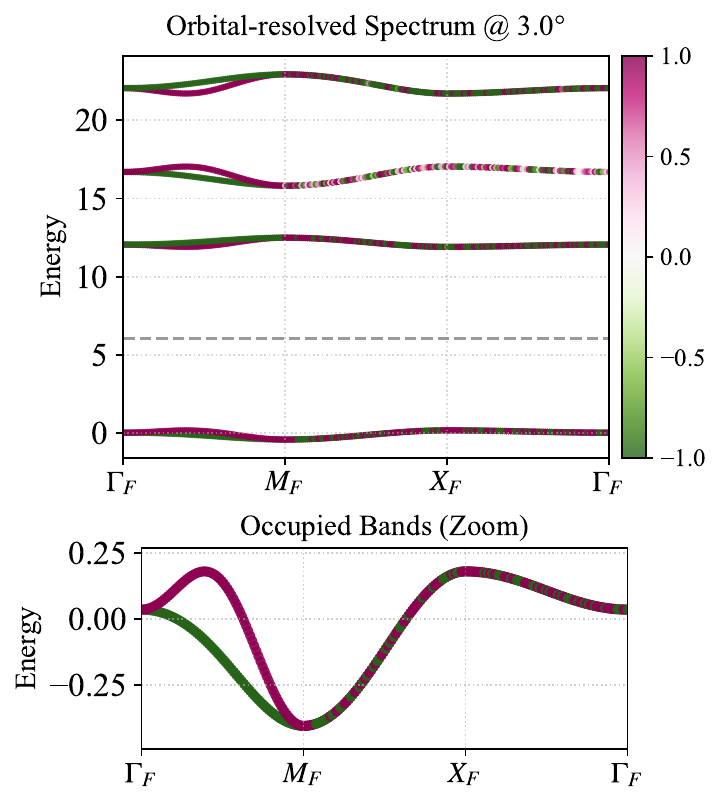}
\caption{Orbital-resolved spectrum at $\theta=3^{\circ}$ in the antiferro-orbital phase. Top: all $8$ bands in the folded Brillouin zone. Dashed line is the chemical potential. Bottom: zoomed in plot of the two filled bands. The color profile corresponds to the difference between the occupations in the two orbitals. One can further perform sublattice projections, and the pink (green) color only appears in sublattice $A (B)$, consistent with the AFO ordering.
}
\label{fig:HF_band_structure_1}
\end{figure}

\begin{figure}[htbp]
\centering
\includegraphics[scale=0.5]{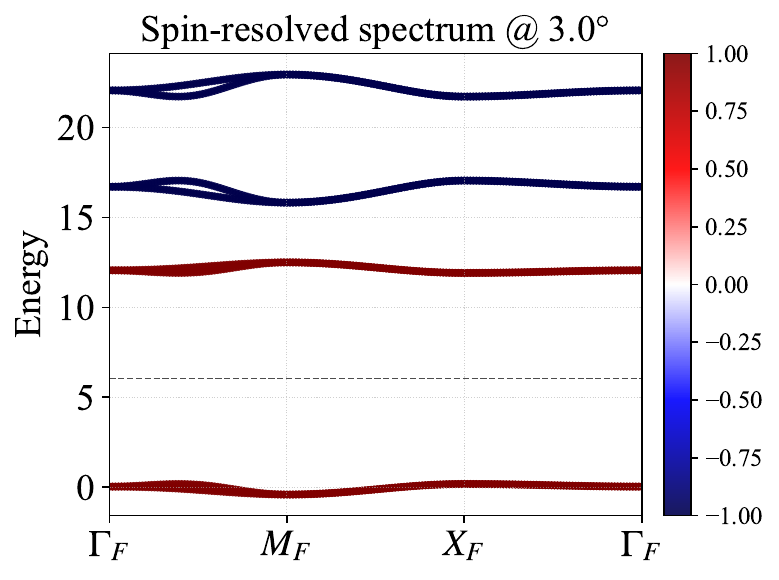}
\caption{Spin resolved band structure at $\theta=3.0^{\circ}$ in the folded moiré Brillouin zone which corresponds to the checkerboard unit cell. The lowest bands are spin polarized.}
\label{fig:HF_band_structure_2}
\end{figure}
From these figures, it's easy to see that there is ferromagnetism in all three phases at zero temperature. While the origin is not transparent to us,  it was proved in ref. \cite{PhysRevLett.112.217201} that when $t_{\pi}=0$, in the $U_{p_xp_x}(0)=U_{p_yp_y}(0)\rightarrow \infty$ limit, at a large range of filling factors, as long as there is nonzero Hund's exchange $J$, the ground state exhibits itinerant ferromagnetism. The results (as well as presence of a related antiferro-orbital order) are further confirmed from Quantum Monte Carlo calculation in ref. \cite{PhysRevX.5.021032} where the case of finite $U$ and tiny $t_{\pi}/t_{\sigma}<0.05$ was also discussed ($t_{\pi}$ introduces sign problem and therefore cannot be fully investigated using QMC). In our setup, the ratio $|t_{\pi}/t_{\sigma}|\sim 0.35$ and is therefore beyond the range of existing literature. While unlike the cases in the literature, our antiferro-orbital phase hosts a gap between the filled and empty bands, the system still exhibits ferromagnetism. It is intriguing to carefully examine the relation between this phase and the previous exact results in future investigations. At finite temperature, by comparison, the continuous spin rotation symmetry of the Hamiltonian implies via the Mermin-Wagner theorem that true long-range spin order must melt.

\begin{figure}[htbp]
\centering
\includegraphics[scale=0.5]{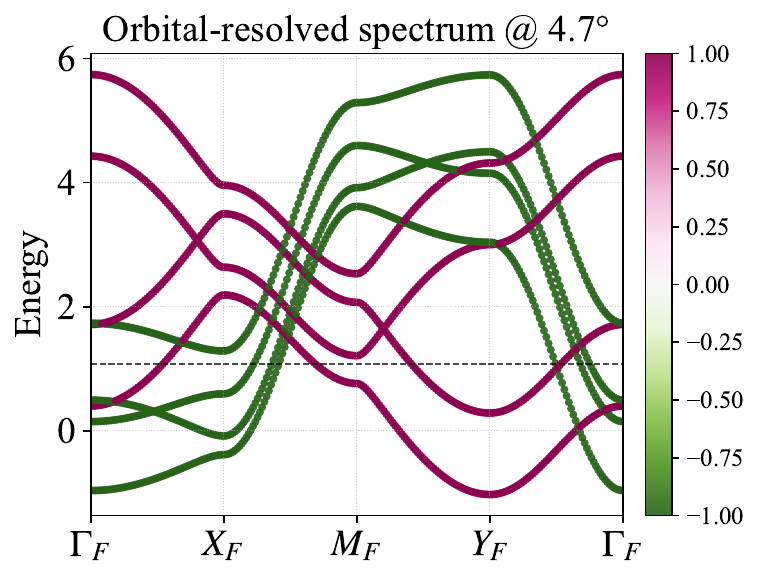}\quad
\includegraphics[scale=0.5]{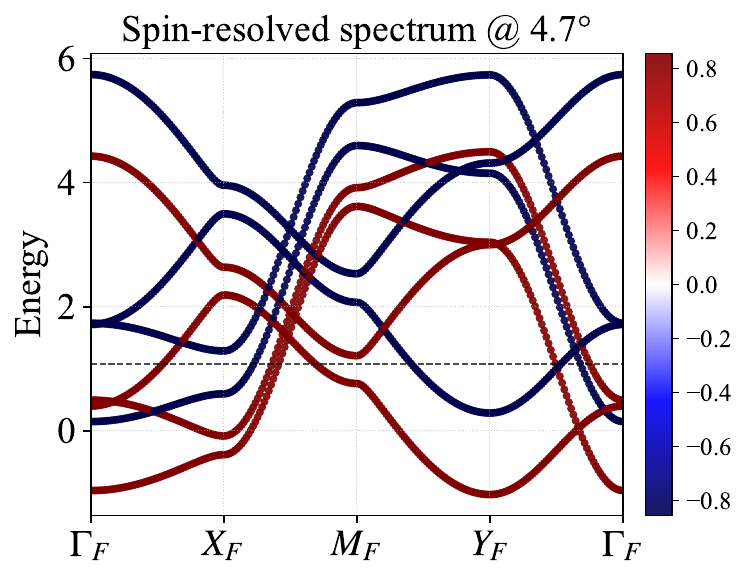}
\caption{Flavor resolved band structures at $\theta=4.7^{\circ}$ in the folded moiré Brillouin zone which corresponds to the stripe geometry.}
\label{fig:HF_band_structure_3}
\end{figure}

\begin{figure}[htbp]
\centering
\includegraphics[scale=0.5]{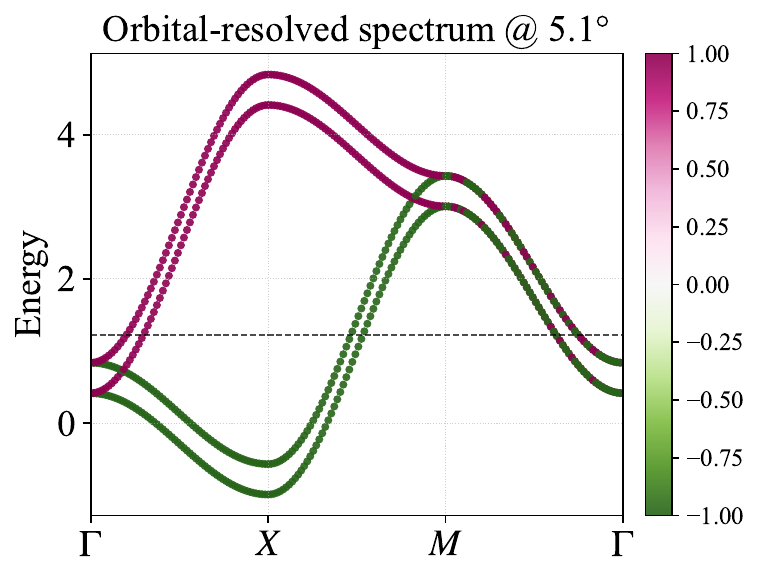}\quad 
\includegraphics[scale=0.5]{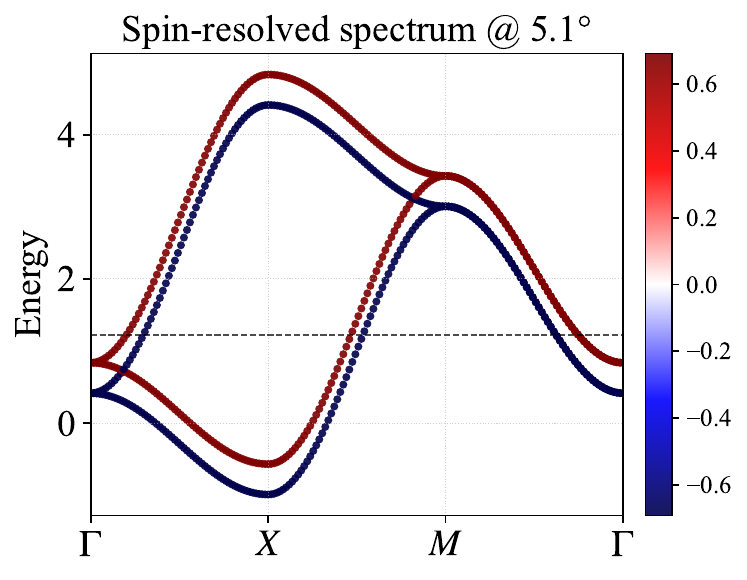}
\caption{Flavor resolved band structures at $\theta=5.1^{\circ}$ in the unfolded moiré Brillouin zone.}
\label{fig:HF_band_structure_4}
\end{figure}

\end{widetext}

\end{document}